%% file: ms.tex
\documentclass[letterpaper,twocolumn,10pt]{article}
\usepackage{usenix2019_v3}

\usepackage{amsmath}
\usepackage{graphicx}

\usepackage{cite}

\usepackage{xspace}
\usepackage{xcolor}
\usepackage{amssymb}
\usepackage{float}
\usepackage{acro}
\usepackage{url}
\usepackage{breakurl}

\usepackage{booktabs}
\usepackage{float}
\usepackage{soul}

\usepackage{multirow}

\usepackage{enumitem}

\usepackage{tikz}

\newcommand{\tikzcircle}[2][red,fill=red]{\tikz[baseline=-0.5ex]\draw[#1,radius=#2] (0,0) circle ;}

\DeclareAcronym{POI}{
	short = POI,
	short-plural-form = POIs,
	long  = point of interest ,
	long-plural-form = points of interest,
	class = abbrev
}

\DeclareAcronym{IOT}{
	short = IoT,
	long  =Internet of Things ,
	class = abbrev
}

\DeclareAcronym{GeoInd}{
	short = GeoInd,
	long  = geo-indistinguishability ,
	class = abbrev
}

\DeclareAcronym{LBS}{
	short = LBS,
	short-plural-form = LBSs,
	long  = location-based service ,
	long-plural-form = location-based services,
	class = abbrev
}

\DeclareAcronym{GP}{
	short = GP,
	long  = Geospatial Programming  ,
	class = abbrev
}

\DeclareAcronym{MCS}{
	short = MCS,
	long  = mobile crowdsourcing  ,
	class = abbrev
}

\DeclareAcronym{LPPM}{
	short = LPPM,
	short-plural-form = LPPMs,
	long  = location privacy-preserving mechanism,
	long-plural-form = location privacy-preserving mechanisms,
	class = abbrev
}

\usepackage{booktabs}
\usepackage{authblk}
\usepackage{cleveref}

\begin{document}
	\date{}
	\title{On (The Lack Of) Location Privacy in Crowdsourcing Applications}%
	\author[1]{Spyros Boukoros}
	\author[2]{Mathias Humbert}
	\author[1,3]{Stefan Katzenbeisser}
	\author[4]{Carmela Troncoso}
	
	\small
	\affil[1]{\textit{Department of Computer Science, TU-Darmstadt, Germany}}
	\affil[2]{\textit{Swiss Data Science Center, ETH Zurich and EPFL, Switzerland}}
	\affil[3]{\textit{Department of Computer Science and Mathematics, University of Passau, Germany}}
	\affil[4]{\textit{SPRING Lab, EPFL, Switzerland}}
	\maketitle
	
	\normalsize
	\begin{abstract}
		\input{Abstract}
	\end{abstract}

	\section{Introduction}
	\input{Intro}

	\section{Mobile Crowdsourcing Applications}
	\label{chapter-DB}
	\input{Datasets}

	\section{Protecting Location Privacy in MCS}
	\label{chapter-defs}
	\input{Defences}
	\input{PrivacyMetrics}

	\section{Existing LPPMs Performance in MCS} 
	\label{sec:empiricaldefence}
	\input{Experiments}

	\section{What's Next?}
	
	\input{Framework}

	\section{Related Work}
	\input{RelatedWork}
	
	\section{Conclusion}
	\input{Conclusion}

	\subsection*{Acknowledgments}
	This work has been funded by the German Science Foundation (DFG) as part
	of the project A1 within the RTG 2050 ``Privacy and Trust for Mobile Users''.

	\bibliographystyle{plainurl}
	\bibliography{refs}

	\appendix
	\section{Appendix}

\input{Appendix}

\end{document}

%% file: Abstract.tex
Crowdsourcing enables application developers to benefit from large and diverse datasets at a low cost. 
Specifically, \ac{MCS} leverages users' devices as sensors to perform geo-located data collection. 
The collection of geo-located data raises serious privacy concerns for users. Yet, despite the 
large research body on \acp{LPPM}, \ac{MCS} developers implement little to no protection for data collection or publication. 
To understand this mismatch, we study the performance of existing \acp{LPPM} on publicly available data from two mobile crowdsourcing projects.
Our results show that well-established defenses are either not applicable or offer little protection in the MCS setting. 
Additionally, they have a much stronger impact on applications' utility  than foreseen in the literature. 
This is because existing \acp{LPPM}, designed with \acp{LBS} in mind, are optimized for utility functions based on users' locations, while \ac{MCS} utility functions depend on the values (e.g., measurements) associated with those locations. We finally outline possible research avenues to facilitate the development of new location privacy solutions that fit the needs of \ac{MCS} so that the increasing number of such applications do not jeopardize their users' privacy.

%% file: Intro.tex

Crowdsourcing is a participative online activity in which the undertaking of a task is outsourced to a group of individuals~\cite{ArolasG12}. This new paradigm of distributing a fragmented task, is an efficient, scalable business model that allows the cheap (or often free) massive collection of data. Indicative of the growth of this data collection methods is the appearance of over 2,000 crowdsourcing platforms\cite{haderer2013apisense,spotteron} in the last years \cite{FungGlobal}. Furthermore, according to recent industrial reports~\cite{Deloitte}, in the last decade, 85\% of top global brands have already adopted crowdsourcing, and in 2018, 75\% of the world's highest performing enterprises would use crowdsourcing. For instance, Google \cite{google}, Microsoft \cite{microsft} and Mozilla \cite{mozillalocation} use crowdsourcing to build WiFi location databases.   

A driving force of the crowdsourcing ecosystem growth is the widespread adoption of smart mobile devices, which enable users to collect geo-located data on their devices and share it with central servers to attain a particular objective. Mobile crowdsourcing applications (MCS) have millions of users around the world. For instance, OpenStreetMaps \cite{osm}, a map generation project from contributed GPS points, reports 4.3 million users in 2018,\footnote{\url{https://wiki.openstreetmap.org/wiki/Stats}} with 1 million active map editors contributing over 4 billion GPS points. Similarly, OpenSignal \cite{opensignal}, a popular network-measuring application, reports over 20 million users.\footnote{\url{https://opensignal.com/methodology\#over-20-million-users-of-our-app}} Safecast \cite{safecast}, a citizen science project collecting environmental data, currently reports over 75 million measurements from approximately three thousand users. 
Many other applications are available \cite{opencellid, radiocells,mozillalocation,skyhook,opensignal,sensorly,cellumap,mapilari,weathersignal, waze,qualcommdrive, kamino,stereopublic}.

\ac{MCS} can bring great benefits for organizations and society. However, the collection and sharing of geo-located data raises serious privacy concerns, as demonstrated by scandals related to the publication of data by fitness applications~\cite{strava,polar} or irresponsible data analysis by transportation companies~\cite{uber-ride-of-glory}. Location data can be used to identify \acp{POI} \cite{krumm2007inference,gambs2014anonymization,freudiger2011evaluating}, infer users preferences, or de-anonymize anonymous traces \cite{zang2011anonymization}. This risk increases when considering auxiliary publicly available information \cite{khazbak2017deanonymizing,olteanu2017quantifying,backes2017walk2friends}, and persists even when protections are put in place~\cite{pyrgelis2017does,DBLP:journals/corr/abs-1708-06145}. 

Over the last decade, the research community has proposed a vast number of \acp{LPPM} to address these issues~\cite{primault2018long}, some of which can provide strong differentially private guarantees \cite{andres2013geo, chatzikokolakis2014predictive, fawaz2014location} and even offer optimal utility \cite{chatzikokolakis2017efficient, oya2017back}. 
Even though it seems like the location privacy question is technically solved, the reality is that these \acp{LPPM} \emph{solely focus on one use case}. They are generally geared towards \acp{LBS} in which users sporadically reveal their location in return for a service (e.g., to find nearby restaurants). In this context, utility is user-centric and hinges on the precision of the reported locations. In \ac{MCS} applications, on the contrary, geo-located data is often shared continuously and over long periods and, while the data utility is still correlated with the location precision, it is foremost tied to the values of the measurements reported at these locations (e.g., WiFi signal strength, or radiation level). Moreover, \ac{MCS} utility cannot be captured with a user-centric approach as, by definition, MCS benefits from aggregating data collected by a large amount of users.

In this paper, we conduct the first in-depth evaluation of the effectiveness of \acp{LPPM} in the context of \ac{MCS}. We use two representative applications, Safecast \cite{safecast} and Radiocells~\cite{radiocells}, which make their contributors' data publicly available on their websites and which have very different utility functions. We propose two new privacy metrics based on statistical measures developed for binary classification and information retrieval to capture the privacy gain provided by the \acp{LPPM} with respect to the identification of areas and points of interest. We also consider new utility measures that, instead of relying on distance-based errors, quantify the accuracy of the aggregate values of data collectively generated.

The results of our experimental evaluation on real data contradict common beliefs regarding the privacy-utility trade-off offered by different \acp{LPPM}. First, location hiding methodologies, which in LBSs help concealing trajectories~\cite{HohGXA07}, do not bring any privacy benefits to \ac{MCS} users. This is mainly because, in \ac{MCS}, the volume of geo-located data is larger and contains points reported over long periods of time (more than a day). Second, differentially private mechanisms~\cite{andres2013geo} offer good protection only for very strong parameters, and even when they are optimized for utility~\cite{chatzikokolakis2017efficient}, they  dramatically perturb the radiation measurements. For instance, in Safecast, we observed that it tremendously changed some areas' radiation levels, urging people to evacuate a place, and completely hindered the ability to localize radiation hotspots (location with elevated radiation). Finally, generalization techniques, usually dismissed in \acp{LBS} because of their poor utility, offer one of the best privacy-utility balance in \ac{MCS}.

In summary, existing \acp{LPPM} are not well aligned with the needs of \ac{MCS} applications. Therefore, new research is needed to approach the design of optimal \acp{LPPM} based on collective, value-based, utility metrics instead of user-centric, location-based utility.

\noindent\textbf{Our contributions} can be summarized as follows:

\noindent{\checkmark} We propose novel privacy and utility metrics suitable to evaluate the performance of \acp{LPPM} for \ac{MCS} data publishing patterns.

\noindent{\checkmark} Using real data collected from two representative \ac{MCS} applications, we show that existing \acp{LPPM} impose too high utility price and that many of them do not even provide good privacy guarantees in the context of \ac{MCS}.

\noindent{\checkmark} We discuss technical and non-technical countermeasures to improve the privacy protection of \ac{MCS} users.

%% file: Datasets.tex

In this section, we introduce the two crowdsourcing applications studied in detail in this paper.

\subsection{Safecast} \label{sec:safecast}
Safecast~\cite{safecast} is a volunteer-centered organization whose goal is to monitor the global radiation levels and detect abnormalities in near real time. Safecast crowdsources the collection of radiation data by providing users with devices that collect radiation measurements every five seconds.

\noindent\textbf{Safecast dataset.}
This dataset contains 64.2 million measurements from 608 users, collected from 2011 to 2017. 
Radiation measurements contain the user's name, a unique user ID, the device's ID, latitude and longitude, a UTC timestamp, and the radiation value and units.
 No registration is required to access these data and Safecast's privacy policy\footnote{\url{https://blog.safecast.org/faq/licenses/}} states that to enable flexibility ``Anyone is free to use with no licensing restrictions''. 
For our experiments we removed IDs corresponding to organizations, malformed entries, and converted all UTC times to local.  
After this process, the dataset has almost 56.7 million measurements from 540 users.

\noindent\textbf{Safecast utility.}
\label{safecast-utility}
The Safecast project uses the collected data to study different phenomena related to radiation. In this paper, we consider two of the main uses of the data.

First, we consider the interactive map to visualize radiation published on Safecast website. Safecast computes the visualized radiation levels from the crowdsourced measurements as follows. 
For a given region of interest, Safecast filters the measurements within the region and computes the average radiation at each location over the last 270 days. 
Second, they discretize the area to 2.25 million grid points (1500 discrete locations per axis). 
They create the displayed map using nearest-neighbor interpolation on the averaged radiation measurements associated to the points of the grid. 
The reported radiation is measured in counts per minute (cpm), expressing how many ionized particles are detected per minute by a monitoring instrument. This use case, which relies on averaging and interpolation, represents a setting in principle amenable to noise in the data.

Second, we consider the detection of \emph{hotspots} -- specific areas where radiation is above a pre-defined threshold. These hotspots indicate locations where radiation could be harmful for public safety. Once identified, Safecast might send experts to perform on-site examination to better understand the causes and consequences of such dangerous zones. Therefore, it is crucial that the localization of hotspots is accurate.

\subsection{Radiocells}\label{sec:radiocells}
Radiocells~\cite{radiocells} is a community project whose goal is to provide an open-source alternative to commercial, closed source, geo-location databases for cell towers and wifi base stations. They also aim to provide raw telecommunication infrastructure data for use in diverse scientific studies. 
Radiocells crowdsources the collection of measurements via a mobile application called `Radiobeacon'.\footnote{\url{https://f-droid.org/packages/org.openbmap/}} 
With this application, users continuously collect measurements as they perform daily activities. Users choose when to start and stop measuring, and when to upload the measurements to the Radiocells server. Furthermore, they can select a specific area where measurements will not be recorded, e.g., to protect their home locations. We do not study the impact of this defense in this paper, but previous work shows that it is rather fragile \cite{hassan2018analysis}. 

\noindent\textbf{Radiocells dataset.} 
The raw data uploaded to the server is publicly available for download. It is licensed under Creative Commons Attribution-ShareAlike 3.0 Unported and ODbL licences aimed at not restricting the use of the data.\footnote{\url{https://radiocells.org/license}}
Amongst other information, the measurements include: signal strength, cell (antenna) ID, location, timestamp, and smartphone model, software, OS version, and manufacturer. In an effort to preserve users' privacy, this dataset does not contain usernames. However, the combination of the smartphone characteristics, the location, and the network provider is likely to represent a quasi-identifier. 
We downloaded data for 2013 to 2017, obtaining 25 million measurements. To separate users' measurements, we grouped the measurements according to phone manufacturer, phone model, country and network operator. 
We obtained 998 potential unique users, of which we only kept those that had more than 100 measurements. We also removed users with spatial inconsistencies, i.e., we removed all users whose speed between two contiguous measurements was greater than 200 km/h. The dataset finally contains 568 users and about 4 million measurements.

\noindent\textbf{Radiocells utility.}
\label{radiocells-utility}
Amongst other purposes, the Radiocells data can be used to geolocate antennas. Such information is useful to enable scientific studies about antennas distribution and signal quality in specific places.
Contrary to Safecast, Radiocells does not provide documentation, nor provide code indicating how they produce their map of antennas. Thus, we use the location function described by OpenCellID \cite{opencellid}, another crowdsourcing project with the same goal, which defines the location of an antenna as the average of the latitudes and longitudes of the measurements referring to this antenna.

%% file: Defences.tex

In this section, we describe the existing \acp{LPPM} we evaluate in our study. These \acp{LPPM} are designed for \acp{LBS} settings, which are different than MCS in two aspects.
First, \acp{LBS} aim at fulfilling an individual need related to one user's location (e.g., find nearby restaurants), while \ac{MCS} aims at fulfilling a common objective through collaborative measurements. Second, \acp{LBS} can often work with sparser geo-located data (just few points per geo-located query) than \ac{MCS}, which requires continuous data collection and in a larger volume.

\subsection{Defenses}
\label{sec:defs}
We consider three type of \acp{LPPM}~\cite{Krumm09,shokri2011quantifying}: (i) spatial obfuscation, (ii) hiding, and (iii) generalization. We do not consider the use of dummy locations or synthetic data~\cite{BindschaedlerS16,ChenAC12}. 
Both approaches focus on producing plausible artificial locations, but to the best of our knowledge there is no proposal that provides the means to generate measurements (or other values) to be associated to these locations. 
In fact, we argue that generating fake measurements, even using prior information, is bound to pollute the real-time measurements that these applications aim at collecting. 

\smallskip\noindent\textbf{Spatial obfuscation.}
The state of the art in spatial obfuscation, which perturbs reported locations with noise, is \textit{\ac{GeoInd}}~\cite{andres2013geo}. This mechanism adapts differential privacy to location data, providing privacy guarantees independent from the adversary's prior information. This approach is widely used in the literature \cite{fawaz2014location,kassem2015anatomization,ma2014nearby,QGIS,pournajaf2015stac,xiao2015protecting}. Following the original definition in~\cite{andres2013geo}, we obfuscate locations by adding planar Laplacian noise. The magnitude of this noise is controlled by the parameter $\epsilon=l/r$ which guarantees that the ratio between the probabilities of two points being the real location in an area of radius $r$ is at most $l$.

\noindent\textit{Release-\ac{GeoInd}.} As with any differentially private mechanisms, in \ac{GeoInd} the level of privacy decreases linearly with the number of reported locations. 
To address this limitation we implement a mechanism inspired by the predictive approach proposed in \cite{chatzikokolakis2014predictive}. This defense reports a new noisy location if, and only if, the user has moved at least $z$ meters away from his previous location. Otherwise, it repeats the last reported location. We call this approach ``Release-\ac{GeoInd}''.

\noindent\textit{\ac{GeoInd}-OR.} Remapping\footnote{A remapping $g$ is a function $g:\mathbb{R}^2\to\mathbb{R}^2$ that maps an output $z\in\mathbb{R}^2$ to another output $z'\in\mathbb{R}^2$ according to the probability density function $g(z'|z)$.} obfuscated locations to popular places according to prior knowledge on users' movements can offer optimal utility without reducing privacy~\cite{chatzikokolakis2017efficient,oya2017back}.  
We  complement \ac{GeoInd} with the remapping approach in~\cite{chatzikokolakis2017efficient}. We refer to this approach as ``\ac{GeoInd}-OR''. 

\smallskip\noindent\textbf{Hiding.}
This defense achieves privacy by suppressing some of the users' locations~\cite{HohGXA07,HuangYMS06}. The released locations are \emph{not} perturbed. We consider two hiding strategies: (i) a ``Random'' strategy in which users release a random subset of their points, and (ii) a ``Release'' strategy in which users only reveal a new point when they have traveled at least $x$ meters away from the previously reported location.

\smallskip\noindent\textbf{Generalization.}
This defense reduces the precision with which locations are reported~\cite{BambaLPW08,GruteserG03}. We implement this approach by reducing the precision of the reported GPS coordinates~\cite{Krumm09}. We denote this defense as ``Rounding''.
%

%% file: PrivacyMetrics.tex

\subsection{Measuring Privacy}
\label{sec:metrics}
Location privacy metrics in the literature are mostly based on a function of the distance between the real location of the user and the one inferred by the adversary \cite{shokri2011quantifying,shokri2009distortion}. This function could measure the \emph{correctness} of the adversary's inference (e.g. using, Hamming or Euclidean distances~\cite{shokri2011quantifying}), or the 
\emph{uncertainty} of the adversary regarding the user's location (e.g., using entropy~\cite{oya2017back}). These metrics are very well suited for the case of \acp{LBS}, where users release one location per query, and the adversary tries to infer that location. However, they are hard to use in the MCS setting, where the adversary has access to locations released continuously over several days. In this case it is hard to establish between which points to compute a distance, or across which points to compute probability distributions for entropy-based metrics.

We also argue that the metrics above do not capture privacy in a manner understandable by users and developers of crowdsourcing applications. How much privacy is an error of 10 meters or 500 meters? It is clear that one is larger than the other, but not how much privacy they provide regarding the potential inference of sensitive information. Even more complicated is the case of entropy, whose units of measurement -- bits, nats, or hartleys -- are rarely known, let alone interpretable, by layman people.

\smallskip\noindent\textbf{Privacy gain.} 
\label{chapter-precisionvsrecall}
We propose to quantify privacy as the loss of adversarial inference power regarding two privacy dimensions understandable by users: geographical area and \acp{POI}. To quantify this loss, we use two
well-established statistical measures: precision and recall. The former captures the increase in privacy when, after a defense, the adversary identifies many false candidate locations along with the user's real whereabouts. Here, the adversary has low \emph{precision} ($\frac{TP}{TP+FP}$, where $TP$ and $FP$ refer to \emph{true positives} and \emph{false positives}, respectively).
The latter captures the increase when, after the defense, the adversary cannot correctly identify the original locations visited by the user. Here, the adversary has low \emph{recall} ($\frac{TP}{TP+FN}$, where $FN$ refers to \emph{false negatives}).

\smallskip\noindent\textit{Spatial privacy gain.} 
Spatial privacy considers the geographical \emph{area} in which the adversary infers the user can be.
We define the true positives ($TP$) as the intersection of the areas where the user can be before and after applying the defense (i.e., the area inferred by the adversary that corresponds to the user's real location). 
Similarly, we define the false positives ($FP$) to be the set difference of the area after the defense and the area before the defense (i.e., the area inferred by the adversary where the user was not present), and false negatives ($FN$) as the set difference of the area before the defense and the area inferred after the defense (i.e., the area where the user has been but that is missed by the adversary).

\smallskip\noindent\textit{\ac{POI} privacy gain.} In reality though, the geographic area itself may not reflect users' privacy~\cite{shokri2009distortion}: if there is only one point of interest in a large area, privacy should be low; and in small areas with many \acp{POI} (e.g., a block in a city), privacy should be large. We propose a complementary metric based on \acp{POI}. In this case, true positives ($TP$) are the \acp{POI} in the intersection of areas before and after the defense is applied. Similarly, false positives ($FP$) are \acp{POI} identified after the defense that were not present before, and false negatives ($FN$) are the \acp{POI} inferred initially that are missed after the defense.

\subsection{Measuring Utility}
\label{sec:utility_meas}
Similarly to privacy, in \acp{LBS}, utility is measured as a function of distance between real and obfuscated locations of one user. This is unsuitable for MCS where location depends on the precision of the aggregate of multiple users' geolocated measurements. We now introduce the utility metrics used in our evaluation.

\smallskip\noindent\textbf{Distance-based.} We call distance-based metrics those associated to LPPMs in the context of \acp{LBS}. In our experiments, we use the per-location haversine distance\footnote{Distance between two points on a sphere given their longitudes and latitudes.} between original and obfuscated locations.

\smallskip\noindent\textbf{Aggregate statistics.}
Most MCS providers are interested in aggregate statistics computed over individuals' contributions. This is the case for Safecast and Radiocells, where the radiation map, respectively the coordinates of the antennas, are derived from average measurements of MCS users. In our evaluation, we consider as MCS utility metrics the actual utility functions of the projects as described in Section \ref{chapter-DB}.

%% file: Experiments.tex

\subsection{Experimental setup}
\label{sec:setup}

We experiment on all data available from Safecast and Radiocells. For Safecast, we additionally consider two regions in Japan with very different radiation profiles: Tokyo, where the radiation profile is quite uniform, and Fukushima, where the nuclear incident at the Daiichi power plant \cite{fukushima} in 2011 created areas with elevated radiation. Table~\ref{tab:datasets-stats} summarizes the statistics (number of users, total amount of measurements, and measurements per user) of the regions under study.

\begin{table}[]
	\centering
	\caption{Safecast (top) and Radiocells (bottom) measurements per region. Vulnerable users are those with at least one cluster.}
	\label{tab:datasets-stats}
	\resizebox{\columnwidth}{!}{%
		
		\begin{tabular}{lccccc}
			\multirow{2}{*}{\textbf{Region}}               & \multirow{2}{*}{\textbf{Users}} & \multirow{2}{*}{\textbf{Measurements}} & \textbf{Average} & \textbf{Standard}  & \textbf{Vulnerable}\\ 
			& & \textbf{} & \textbf{per user} & \textbf{deviation}  & \textbf{users} \\ \cmidrule[1pt]{1-6}
			Tokyo                & 30    & 2,701,367          & 90,046        & 203,576  &  24 (80\%) \\ 
			Fukushima                     & 104   & 7,765,773          & 74,671        & 260,671  &  65 (62\%) \\ 
			World                        & 540    & 56,655,768         & 105,504       & 70,954   &  349 (65\%)  \\ 
			
			\cmidrule[0.5pt]{1-6}
			World     & 568    & 3,710,547    & 6,532       & 17,312 &  91 (16\%) \\
			
	\end{tabular}}
\end{table}

We evaluate the privacy gain and the utility loss of an \ac{LPPM} as follows: 

\noindent\textit{Step 1. Adversary's inference.} Inspired by previous works, we use clustering to implement inference
on the regions and the points of interest for all users.~\cite{cho2016exploiting,drakonakis2019please,mathew2012predicting,krumm2007inference,hoh2006enhancing,freudiger2011evaluating,urnerassessing,ester1996density}. Concretely, we use the density-based clustering algorithm (DBSCAN)~\cite{ester1996density}. Contrary to other clustering algorithms (such as K-Means), DBSCAN is robust to noise and outliers and does not require to specify the number of clusters a priori (see Appendix~\ref{ap:dbscan}). We keep the five clusters with the highest number of points, and \emph{we consider their total area as the geographical area input to the Spatial Gain metric}.
Once clusters are identified, we use the OSM API\footnote{\url{https://wiki.openstreetmap.org/wiki/Overpass\_API}} to find the \acp{POI} in the clusters of the targeted user. \emph{We consider all points in the top five clusters as input for the POI Gain metric}.  
Table~\ref{tab:datasets-stats} reports the percentage of users \emph{vulnerable} to our attacks \emph{before} the defenses are applied, i.e., the percentage of users for which we find at least one cluster. For Safecast-Tokyo, we only report statistics for the 30 users considered when using GeoInd-OR (see Section \ref{sec-prv-gain}).

\noindent\textit{Step 2. LPPM Application.} We apply the \ac{LPPM} to all users' data and repeat the actions in Step 1 to infer their regions and points of interest. 
Note that when Rounding to 2 or 3 decimals, obfuscated locations are separated by approximately 1,100 meters and 110 meters, respectively. Thus, our parametrization of DBSCAN is bound to not find any clusters. However, an adversary would know that given an obfuscated point, the actual location of the user is within a square of size 110, resp. 1,100 meters, centered in the reported location. Thus, for this case, instead of using DBSCAN clusters, we pick the squares of the respective sizes around the five most frequently reported obfuscated locations.

\noindent\textit{Step 3. Privacy gain.} We compare the area (in square kilometers) of the clusters before and after the \ac{LPPM} to compute the Spatial privacy gain, and the \acp{POI} inside the clusters for the \ac{POI} privacy gain.

\noindent\textit{Step 4. Utility loss.} In the case of aggregate statistics, the utility loss is application dependent. For Safecast, we consider the absolute difference in cpm per grid point between the radiation values on the application's interactive map (see Section~\ref{sec:safecast}), before and after the \ac{LPPM}. In Radiocells, we consider as utility loss the distance between the location of the antennas before and after the \ac{LPPM}.

\input{threat}

\subsection{Privacy Gain}
\label{sec-prv-gain}

\begin{figure*}[t]

  \centering
    \includegraphics[width=0.8\textwidth]{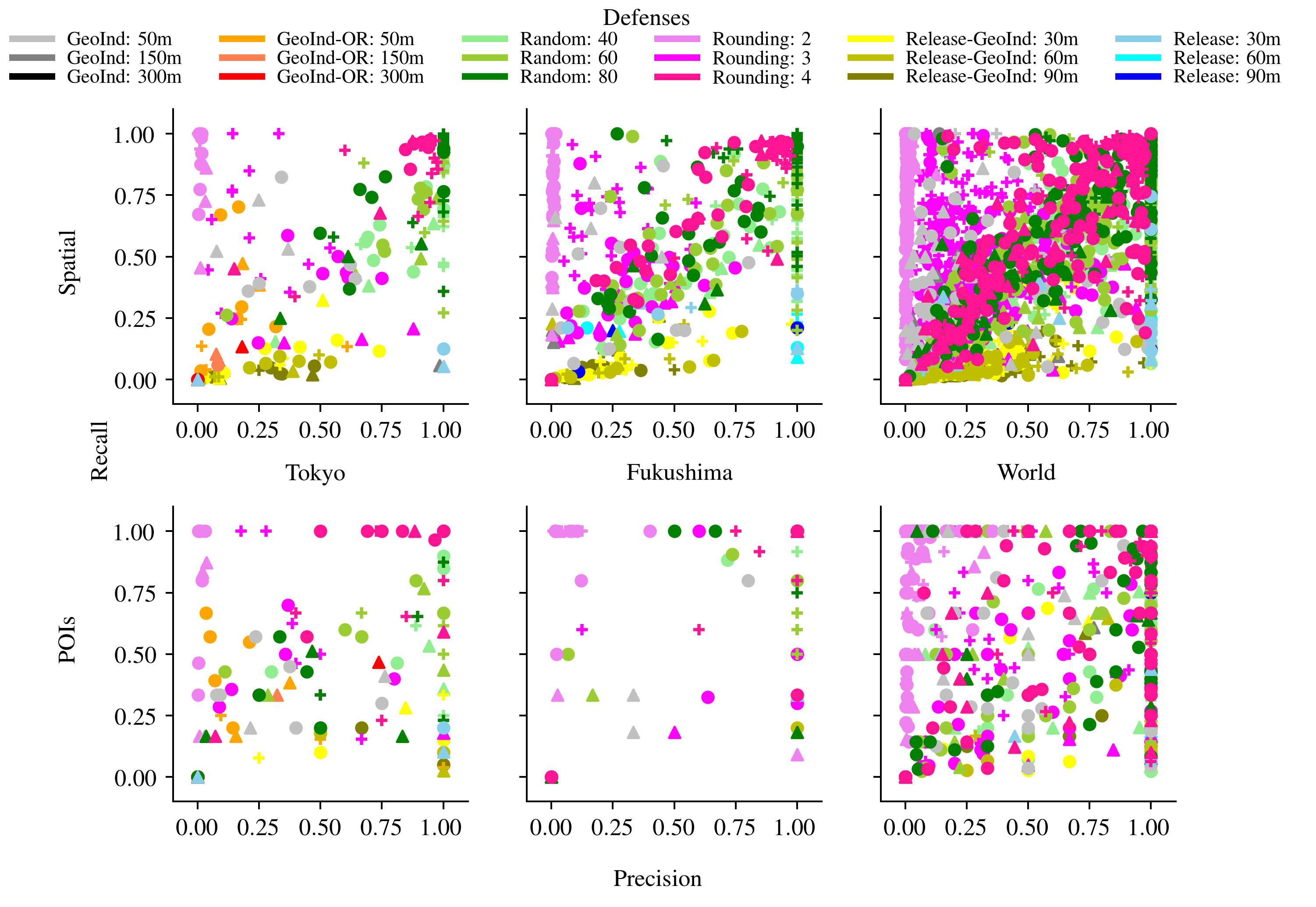}
     \caption{Safecast privacy gain: Spatial (top) and \acp{POI} (bottom).
     Amount of measurements per user  \textbf{+} : \textless{}10k, \tikzcircle[black, fill=black]{2pt} : [10k,50k], 
     $\blacktriangle$ : \textgreater{}50k. Each point on the graphs represents one user.}
      \label{figure:clusters-overlap}
\end{figure*}

\noindent\textbf{Defenses implementation.}
For the \ac{GeoInd} defense, we set the privacy parameter $l=\ln(1.6)$, and use radius $r\in\{50, 150, 300\}$ meters which yields $\epsilon \in \{0.01,0.003,0.001\}$. Remapping the locations for the \ac{LPPM} GeoInd-OR requires computing the posterior probability for every candidate location. This operation is rather costly when the number of locations being considered grows. To keep a reasonable experimentation time, we only test GeoInd-OR for the Tokyo region in the Safecast dataset. We use 80\% of the users to construct the prior probability distribution describing users’ movements, and the remaining 20\% to evaluate the effectiveness of the approach. We chose this 20\% manually to keep a balanced testing set. It is composed of the top 10 users with many (more than 50k), moderate (between 10k and 50k), and few (less than 10k) measurements. Finally, for the Release-GeoInd mechanism, we use $l=\ln(1.6)$, $r$ = 50 meters, and we select the distance between released locations to be $z\in\{30,60,90\}$ meters. We provide details about the implementation of these \acp{LPPM} in Appendices \ref{ap:geoind} and \ref{ap:OR}.

We implement the Random mechanism tossing a biased coin every time a location is about to be reported. The bias is set so that users release on average 40\%, 60\% or 80\% of their measurements.
For the Release mechanism, we sort all the locations reported by a user in chronogical order, and release a new location only if it is separated by (at least) $x\in\{30,60,90\}$ meters from the previously reported one. If two locations are less than $x$ meters apart but in different days, we release them both. 

Last, we implement Rounding by rounding to 2, 3, or 4 decimals the latitude and longitude of the users' locations. Effectively, this reduces the location accuracy to roughly 1,100 meters, 110 meters and 11 meters, respectively.

\subsubsection{Safecast}
We first evaluate the privacy gain of the \acp{LPPM} in the Safecast dataset. Figure \ref{figure:clusters-overlap} shows the Spatial (top) and \ac{POI} (bottom) gain for Tokyo, Fukushima, and the whole world.
(Figure~\ref{figure:world-geoind} in the appendix shows the results for each of the defenses separately for the whole world.)
The x-axis represents precision, and the y-axis recall. Each point in the graph represents a user, and the markers' shape indicate the amount of measurements she contributes. The colors represent the \acp{LPPM}.  To compute these graphs, we configure DBSCAN to find clusters with at least 75 points separated by at most 30 meters (roughly the size of a small building). As in~\cite{drakonakis2019please}, we also require that, for each cluster, users either stay more than 30 minutes, or visit it for at least two days. A first reason to fix these parameters is to evaluate the gain for all users under the same conditions. A second reason is that the loose parameters used in Section~\ref{sec:attacks} can yield very sparse clusters with few points that are hard to break by removing or perturbing locations. Thus, the defenses would perform equally bad and we would gain little information about their properties. Tightening the parameters reduces the work inference success to 21\% (some clusters are not found), which still represents a significant risk.

Defenses that provide large gains result in points close to the figure axes. Points near the y-axis indicate low precision, i.e., cases in which the adversary correctly identifies some (or even all) of the true locations but also inferred many other wrong locations. Points near the x-axis indicate low recall, i.e., cases in which the adversary correctly identifies some real locations, but misses many others.
Unsurprisingly, we observe a high variance in the defenses' performance since it is highly dependent on the user behavior. However, it is possible to identify some trends. 

We first discuss the Spatial privacy gain (Figure \ref{figure:clusters-overlap}, top). For the least privacy-preserving parameter ($r=50$m), GeoInd significantly decreases the number of vulnerable users (grey points in the figure) from the values reported in Table~\ref{tab:datasets-stats}. The reduction is 50\% for Tokyo (from 24 vulnerable users to 12), 45\% for Fukushima, and 45\% for the whole world. When the mechanism is strengthened ($r=300$m), GeoInd adds so much noise (see Figure \ref{figure:geoind-cdf} in Appendix~\ref{DefensesSubsectionAppendix} for reference) that no users are vulnerable after the defense. 
In summary, GeoInd seems to provide fairly good privacy gain in Tokyo and Fukushima. Yet, when we look at the whole dataset, it becomes clear that the protection provided by \ac{GeoInd} is highly dependent on the users' movement patterns. 

The Release-GeoInd (yellow) mechanism works generally better than \ac{GeoInd}. Even though more users are vulnerable (only between 4\% and 13\% of the users become not-vulnerable) and the adversary obtains reasonable precision, it yields very low recall. This is because in this method users keep reporting the same obfuscated location until they move. This repetition results in clusters being found on fake locations that often do not overlap with the original ones. This reduction becomes more significant as the defense is configured to provide more privacy (larger $z$).%

GeoInd-OR performs slightly better than vanilla \ac{GeoInd}. This is because the remapping results in points being repeatedly mapped to popular places causing the generation of clusters around those not-real locations. 

Similar to vanilla \ac{GeoInd}, the Release mechanism (blue) significantly reduces the number of vulnerable users -- by more than 50\% even for the least conservative parameter.
However, when precision is very high, i.e., when a cluster is found, it corresponds to a real location. The reason is that even though the user hides many points, if a location is visited regularly, the user will eventually report enough points around this location to make the cluster identifiable by the adversary.

The Random hiding mechanism (green) does not perform well. First, it reduces the number of vulnerable users less than other defenses (10\% decrease in Tokyo, 27\% in Fukushima, and only 5\% when considering the whole world). From the vulnerable users only a handful obtain good protection. We could not find a clear pattern to predict which movement profiles would best benefit from this defense. For many users, especially those with a few points, removing points at random still yields high precision as the few measurements are very localized. Overall, we do not notice much influence of the fraction of hidden points on the privacy of the users.

Finally, the protection provided by Rounding (pink) depends on the rounding parameter. Keeping 4 decimals reduces accuracy by just 11 meters. Therefore, the adversary finds roughly the same clusters, i.e., for many users we observe high recall and precision after the defense (especially in Tokyo and Fukushima). On the contrary, rounding to 2 or 3 decimals significantly increases the size of inferred spatial areas, which leads to variable recall (depending on the users' movement patterns) and low precision. 

Regarding the \ac{POI} privacy gain (Figure \ref{figure:clusters-overlap}, bottom), a first observation is that the amount of users vulnerable to the attack, i.e., points in the graph, is lower. This is because for many users the identified clusters do not contain any \ac{POI} (according to the OSM API).
Second, for the users who have \acp{POI} in their clusters, both recall and precision are higher than in the Spatial gain. This is because many of the large clusters that contribute to the low Spatial precision do not have \acp{POI} and thus do not contribute to the confusion of the adversary when identifying particular POIs. Furthermore, the clusters that the adversary finds after the \acp{LPPM} may cover a smaller area than the original clusters, but still contain most of the users' initial \acp{POI}. This provides a higher \ac{POI} recall than  Spatial recall.
Third, in this case we observe a significant difference between Tokyo and Fukushima. The reason is twofold. First, the Fukushima prefecture is much larger than the area of Tokyo we consider. Second, Fukushima is a rural area and thus contains fewer \acp{POI} than Tokyo where even small clusters have many places of interest. 

These observations reinforce previous insights that solely considering the spatial dimension may provide a false perception of privacy~\cite{shokri2009distortion}. Considering a \ac{POI}-privacy measure is necessary for providing a comprehensive picture of the privacy threat users face in \ac{MCS} applications. We note that this perception also depends on DBSCAN parameters, which define the size of the regions found, and consequently the number of POIs, increasing the manual effort of the adversary. We discuss this effect in Appendix~\ref{Exp_res_appendix}.

\smallskip\noindent\textbf{\textit{Impact of the amount of measurements on privacy.}}
\label{points-impact}

\begin{figure}
  \centering
    \includegraphics[width=0.9\columnwidth]{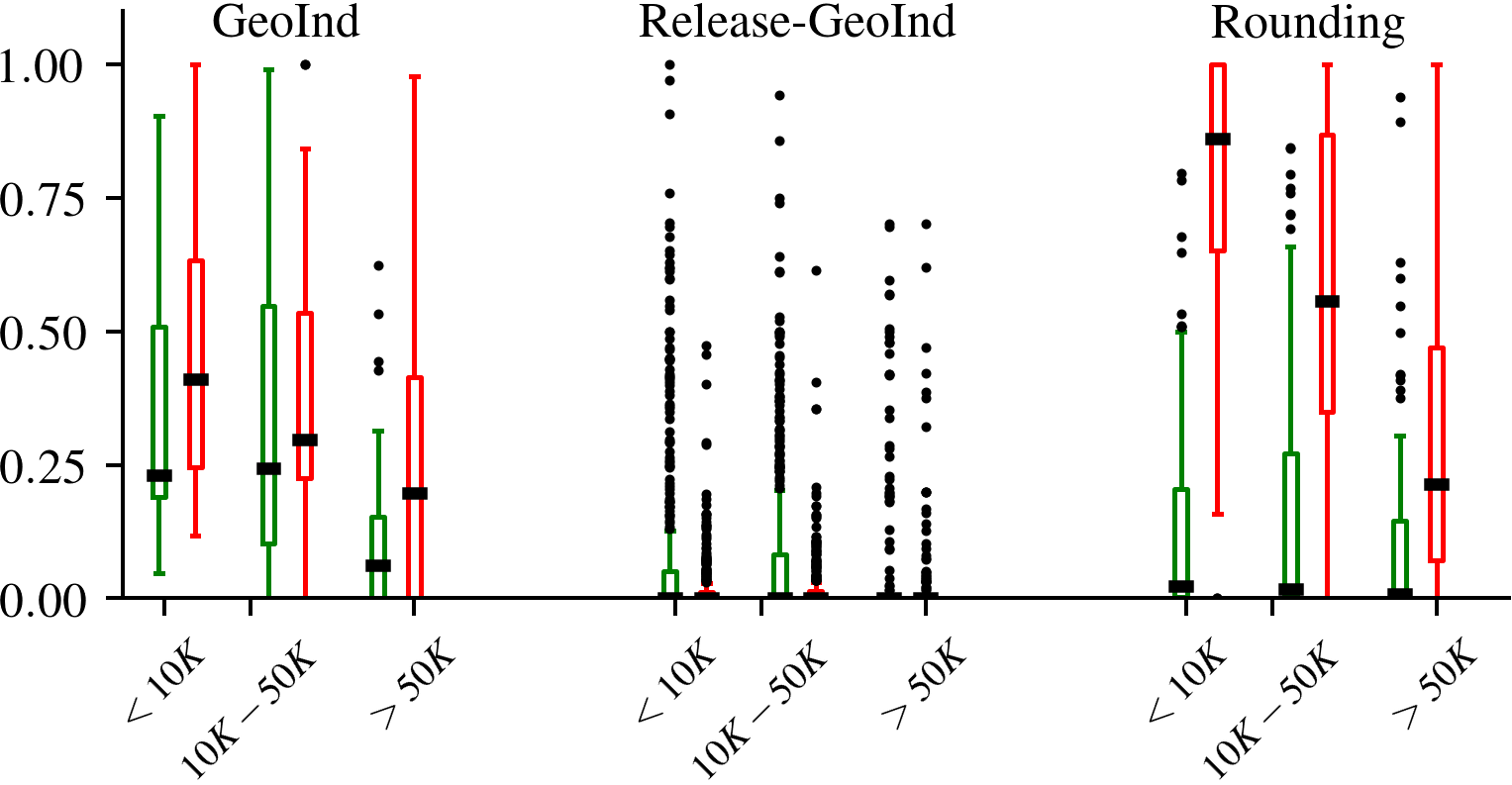}
          \caption{Precision (green) and recall (red), depending on the amount of measurements $x$ per user for three selected defenses (all parameters combined).}
             \label{fig:permeasurements}
\end{figure}

We present in Figure \ref{fig:permeasurements} the Spatial gain for the three best \acp{LPPM} (all parameters combined) split by the amount of measurements users contributed.
We discard Rounding 4 as it does not provide any privacy. We see that all \acp{LPPM} provide low precision and recall regardless of the users' contribution volume. The exception is Rounding which, as explained above, by definition provides variable recall and low precision.

Counterintuitively, the \acp{LPPM} perform worse for users who contribute fewer points. This is because the attack constructs more, and larger (on average 10 times bigger), clusters for people who share many points than for those sharing fewer points. These clusters are split after the \acp{LPPM} are put in place, as some reported locations are moved away from their original clusters while other measurements, perturbed with noise, concentrate to new places forming wrong clusters. For Rounding, where every cluster created after the \ac{LPPM} has roughly the same size, users with a few measurements have higher recall because their initial small clusters are often covered by the large regions resulting from the \ac{LPPM}.

\smallskip\noindent\textbf{\textit{Thwarting workplace inference.}}
Finally, we evaluate the effectiveness of the different \acp{LPPM} at hiding workplaces. 
Recall that, without protection, we can identify the workplace of 21\% (7 out of 34) of the users. 
Five defenses, GeoInd, Release-GeoInd, Release, and Rounding 2, protect all users from inferences. Random hiding requires heavy sampling to be effective (hiding only 20\% permits the identification of 6 workplaces, and hiding 40\% still reveals 1). Finally, unsurprisingly, Rounding to 4 decimals does not protect against work inference, and Rounding with 3 decimals only hides one workplace out of 7.

\begin{figure}[ht]
	\hspace{-0.8cm}
	\includegraphics[width=1.15\columnwidth]{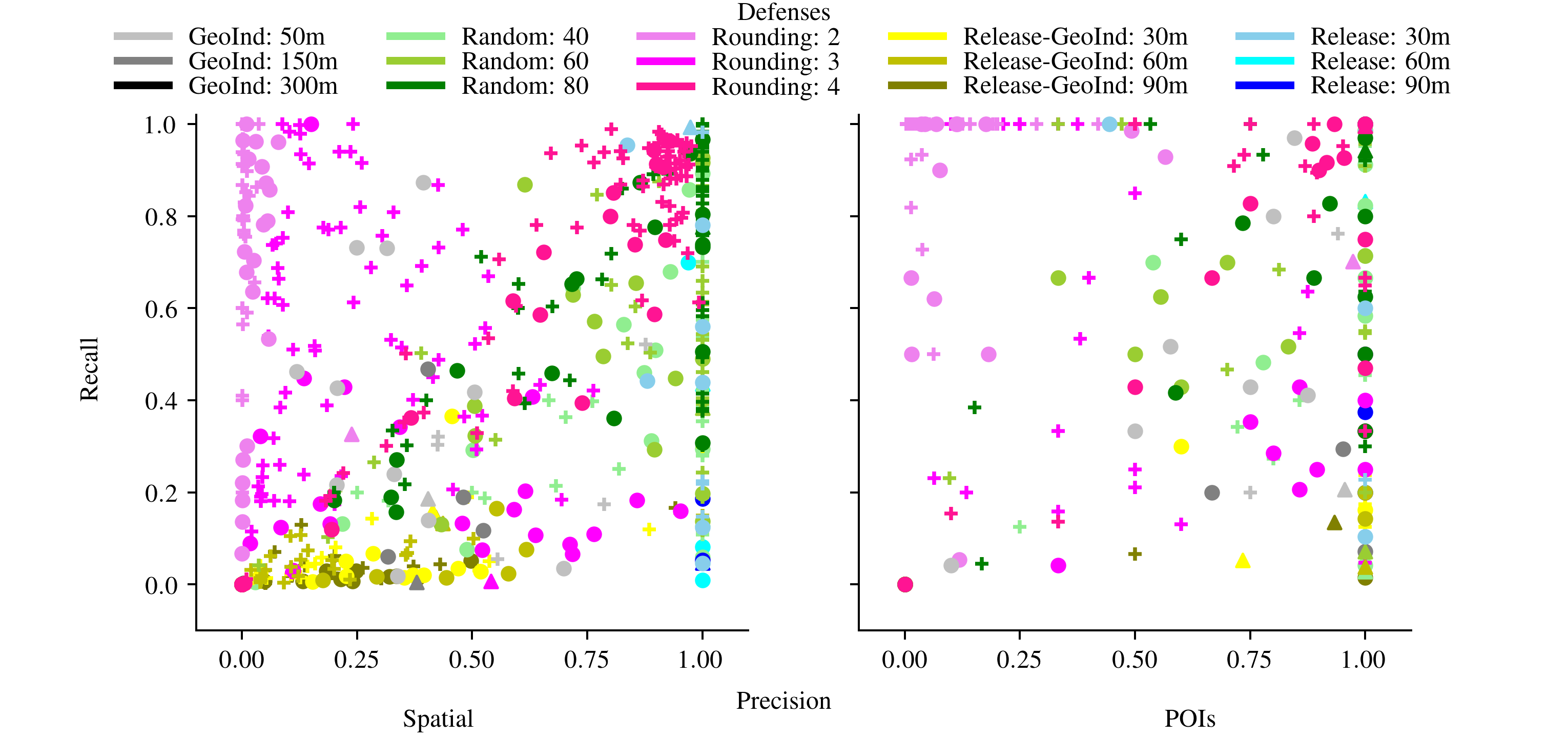}
	\caption{Spatial privacy gain (left part) and \ac{POI} privacy gain (right part) in Radiocells.
		Amount of measurements per user \textbf{+} : \textless{}10k, \tikzcircle[black, fill=black]{2pt} : [10k,50k], 
		$\blacktriangle$ : \textgreater{}50k. Each point on the graphs represents one user.}
	\label{figure:radiocells-clusters}
	
\end{figure}

\subsubsection{Radiocells}
Users in Radiocells have on average fewer measurements than those in Safecast, and clustering requiring 75 points yields very few clusters. Hence, for this dataset we loosened the DBSCAN requirement to 25 points per cluster.

We see in Figure \ref{figure:radiocells-clusters} that \ac{GeoInd}-based mechanisms behave similarly to the Safecast case in terms of Spatial gain: GeoInd provides highly variable protection, and Release-GeoInd yields low recall while precision depends on the user behavior. Vanilla \ac{GeoInd} decreases the number of vulnerable users by 14\%, and Release-GeoInd by 2\%. Given that only 16\% of the users were initially vulnerable, this reduction is significant.
For the hiding mechanisms, the Random and the Release mechanisms decrease the number of vulnerable users by 7\% and 14\%, respectively. For the vulnerable users, contrary to Safecast, these mechanisms consistently yield high precision, i.e., they offer poor privacy protection for Radiocell's users movement profiles. Finally, the Rounding mechanisms with parameters 2 and 3 offer reasonable privacy. Regarding \acp{POI}, we observe similar behavior to the Safecast dataset.

Overall, the results in Radiocells are consistent with our findings in the Safecast dataset, confirming the trends regarding the \acp{LPPM} behavior in the \ac{MCS} setting.

\subsection{Privacy-Utility Trade-Off}
\label{sec:utility}

\subsubsection{Safecast} 

\begin{figure}
	\centering
	\includegraphics[width=0.95\columnwidth]{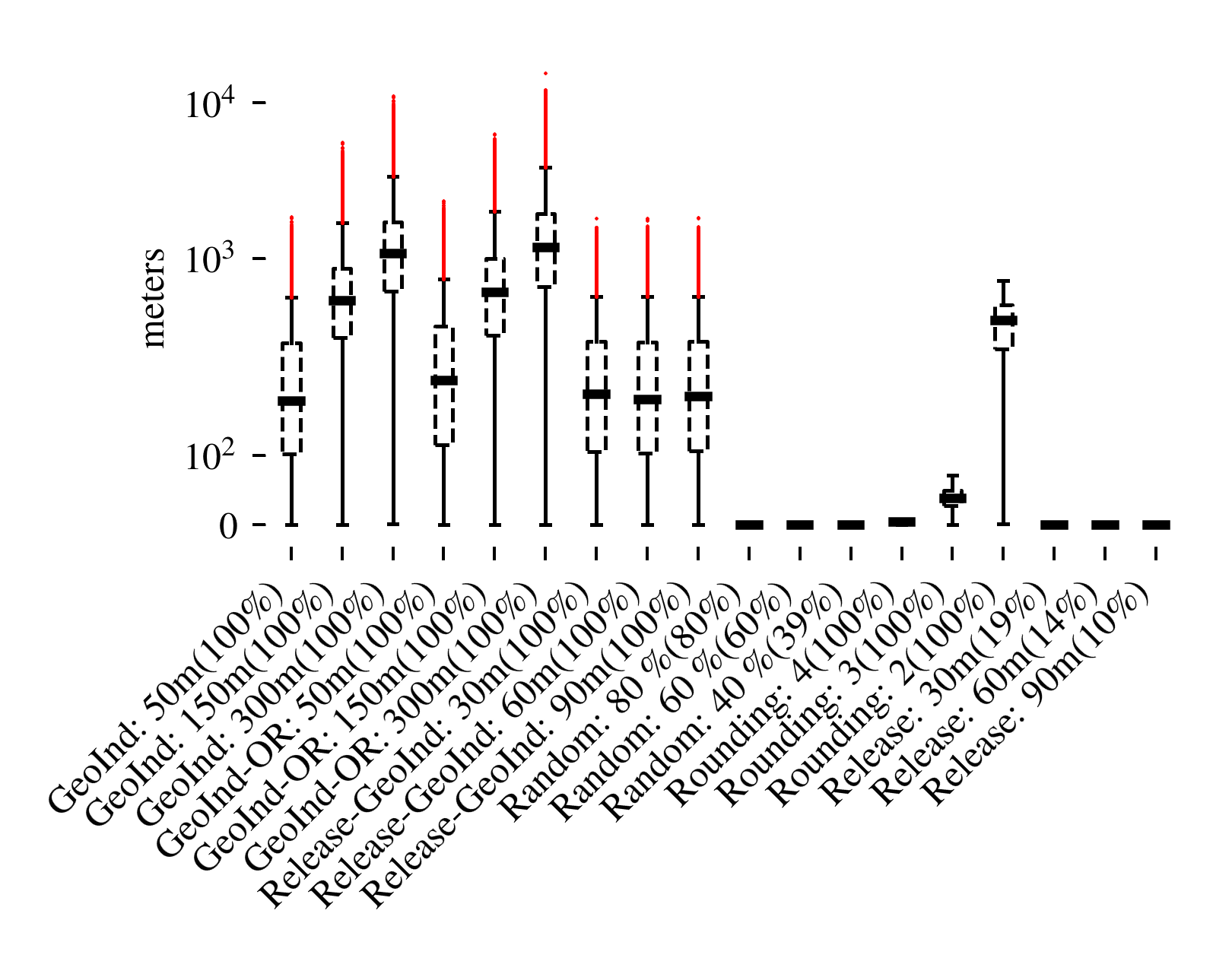}
	
	\caption{Measurement error in Tokyo using a distance-based metric. This can be interpreted either as privacy gain or utility loss.}
	\label{fig:utility-LBS-Tokyo}
\end{figure}

\smallskip\noindent\textbf{\textit{Distance-based metric vs Aggregate statistics for \ac{MCS}.}}
\label{LBS-utility}
\label{agg_stats}
We first evaluate the utility loss incurred by the \acp{LPPM} measured using the LBS-oriented distance-based metric described in Section~\ref{sec:utility_meas}. This utility metric is based on the distance between reported and real locations, but disregards the (radiation) values that Safecast cares about.
Figure \ref{fig:utility-LBS-Tokyo} displays the results for users in the Safecast-Tokyo dataset. The y-axis indicates the distance in meters, and the x-axis the \ac{LPPM} and the percentage of points that are released. Random and Release \acp{LPPM}, which add no noise, are the best in terms of error; and GeoInd \acp{LPPM} offer the worst performance as they tend to spread locations --- sometimes more than a kilometer away from the initial measurements (see Figure~\ref{figure:geoind-cdf} in Appendix~\ref{DefensesSubsectionAppendix}).

\begin{figure}
	\centering
	\includegraphics[width=0.95\columnwidth]{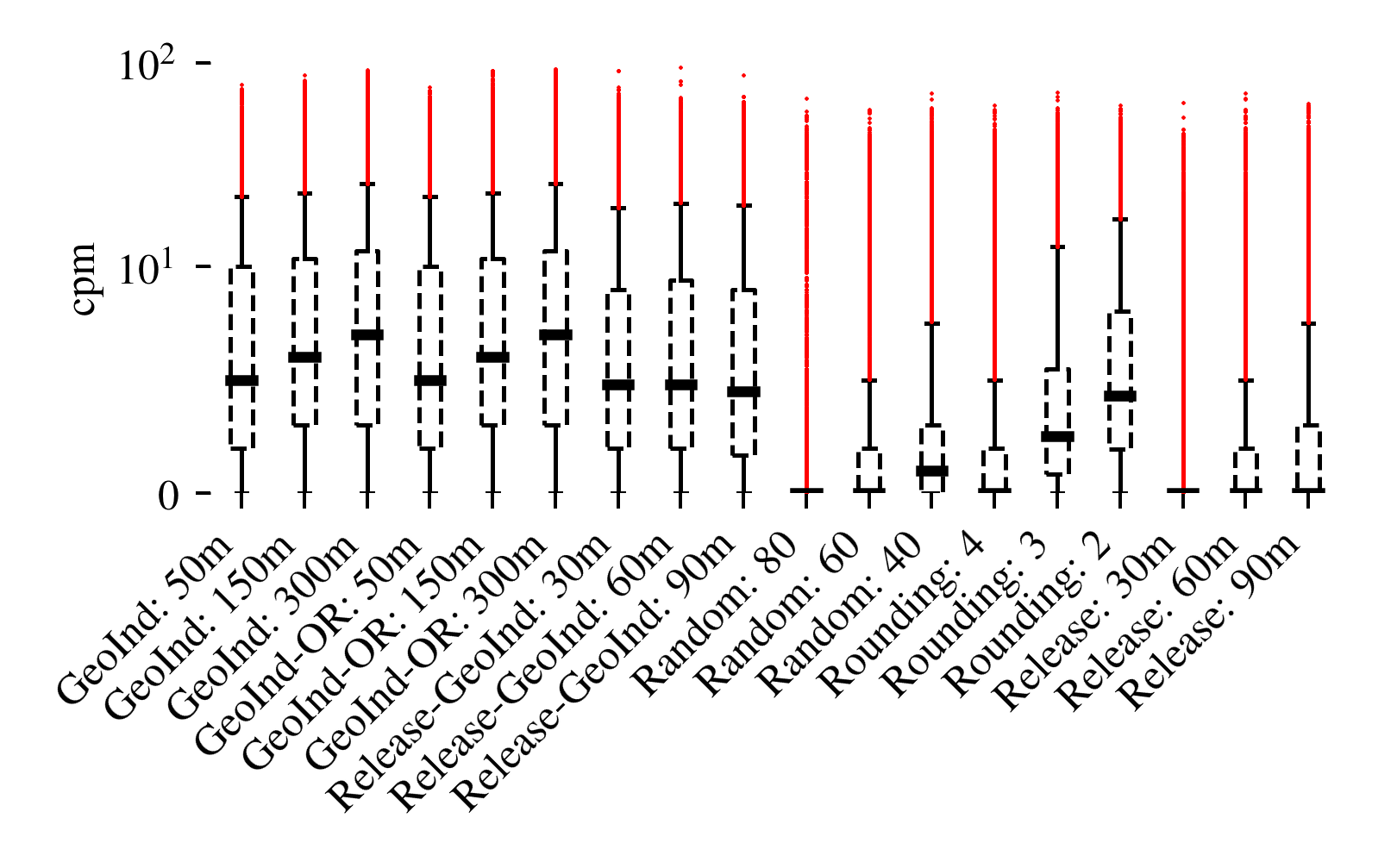}
	
	\caption{Absolute difference in Tokyo's radiation values with Safecast dataset.}
	\label{fig:utility-tokyodif}
\end{figure}

\begin{figure}
	\centering
	\includegraphics[width=0.9\columnwidth]{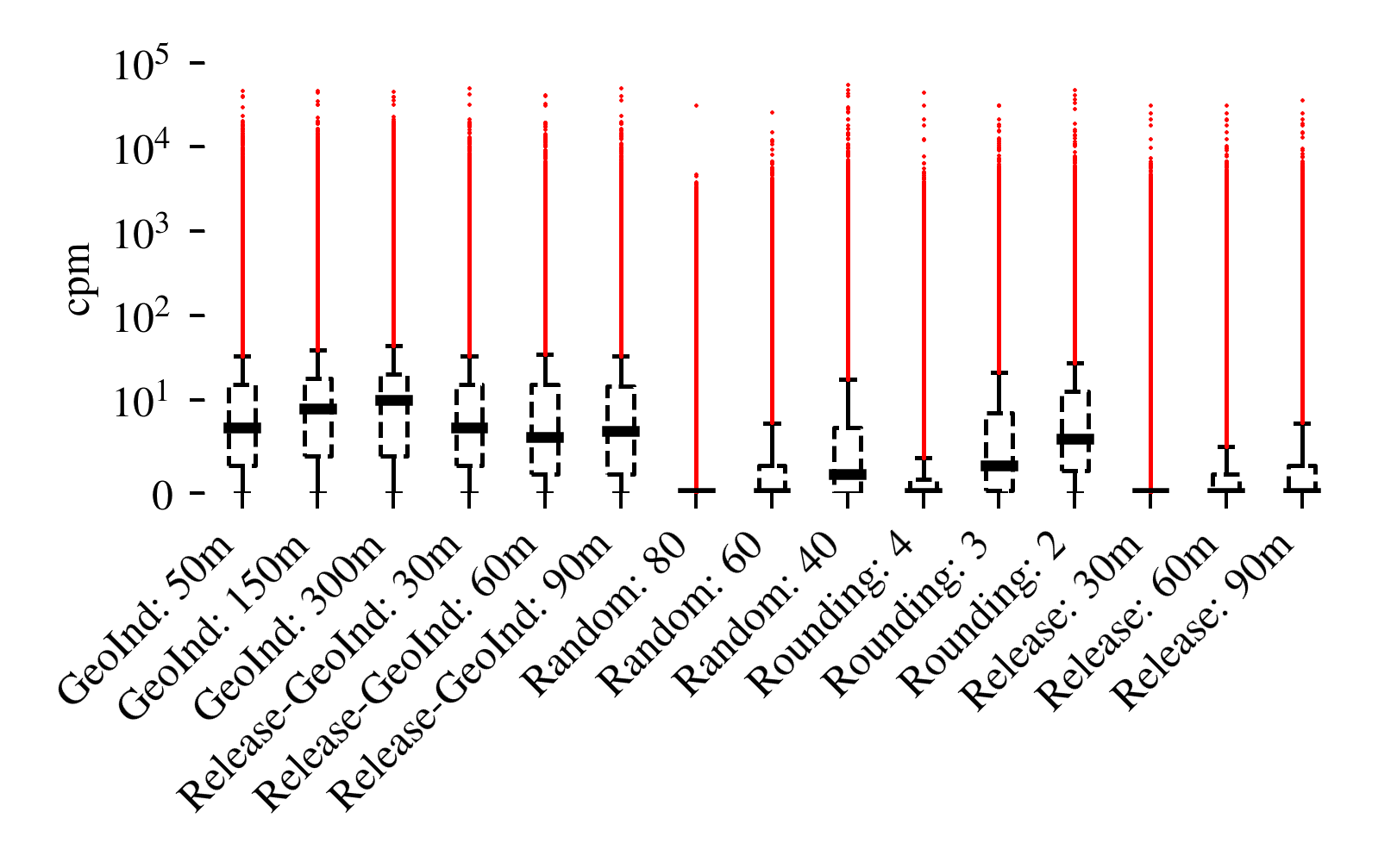}
	\caption{Absolute difference in Fukushima's radiation values with the Safecast dataset.}
	\label{fig:utility-fukushimadif}
\end{figure}

Next we consider the utility loss for aggregated statistics, i.e., utility measured as the difference between radiation values to be plotted on the generated map. We plot per grid-point utility loss for Tokyo and Fukushima in Figures \ref{fig:utility-tokyodif} and \ref{fig:utility-fukushimadif}, respectively. We observe that the loss is similar in both regions, though in Fukushima the median loss is slightly higher and there are more, and larger (up to $10^{4}$ radiation offset with respect to the original value), outliers than Tokyo.  
Because of the interpolation step, in this case all GeoInd variants offer roughly the same utility loss on average. Still, Hiding and Rounding strategies offer better performance, with small median error for the least protective parameters.

If we compare the distance-based results (Figure \ref{fig:utility-LBS-Tokyo}) to the aggregated statistics utility loss (Figure \ref{fig:utility-tokyodif}), we observe significant differences. 
First, the interpolation step results in \acp{LPPM} based on \ac{GeoInd} to fare much better in terms of aggregates than in terms of distance. 
Second, distance-based metrics underestimate the utility loss of hiding \acp{LPPM} (Random and Release). While it is true that the released points have no error in distance, hiding points comes at a cost not reflected in the metric. This is made evident by the aggregated metrics, which show that the more points are hidden, the larger is the utility loss. We note that relying on Markov mobility models such as in~\cite{gambs2012next,shokri2011quantifying} could help interpolate the hidden locations. Yet, this would not help recover the (radiation) values attached to them and the utility loss would remain. For the generalization mechanisms, distance-based metrics consistently report larger median loss, but have less variance and less outliers.  

In summary, distance-based metrics provide a very different perception of the \ac{LPPM} performance than considering utility functions computed on the geo-located values, overestimating the performance of some methods (e.g., hiding strategies) and underestimating others (e.g., GeoInd-based \acp{LPPM}). We conclude that traditional LBS-oriented metrics are inadequate for measuring utility in \ac{MCS} scenarios.

\smallskip\noindent\textbf{\textit{Semantic interpretation.}}
The absolute difference in cpm of measurements before and after the defense gives a rough idea about the utility loss, but it is difficult to interpret. Is it significant? What is the effect of outliers? Does reporting the values after the defense have any implication on the danger for human health? To answer these questions, we study how the variance introduced by the defenses can change the interpretation of the risk at a given location. To this end, we rely on the cpm safety scale \cite{geiger} provided with one of the top-seller Geiger counters (radiation measurement devices) on the market. This scale contains five categories:
\begin{itemize}[noitemsep,topsep=0pt]
	\item Category 1: 0-50 cpm. Normal radiation background.
	\item Category 2: 51-99 cpm. Medium level. 
	\item Category 3: $>$100 cpm. High level.
	\item Category 4: $>$1000 cpm. Very high level, leave area.
	\item Category 5: $>$2000 cpm. Extremely high level, immediate evacuation.
\end{itemize}
We select the prefecture of Fukushima and two defenses that produce a good level of privacy: \ac{GeoInd} 300m and Rounding 2.
For each of the 2.25 million grid-points on Safecast's radiation map for Fukushima, we compute their radiation category according to the safety scale before and after each defense. For \ac{GeoInd} 300m, which is of probabilistic nature, we repeat the procedure 10 times and report the average. We present the results in Tables \ref{table:fuku-geoind300} and \ref{table:fuku-round2}. 
We observe that the majority of the points either stay in their original category or move to a nearby. However, we observe some extreme category jumps from the first category (safe radiation levels) to the fourth and  fifth (high danger). 
For instance, \ac{GeoInd} causes 53 places to be marked as dangerous instead of safe. Even more alarming, 283 locations that should be marked as extremely dangerous are marked as safe or slightly elevated (categories 1 and 2).
On the contrary, the Rounding mechanism limits the number of extreme changes. For instance, there is a category jump from 5 to 1 and 2 only for 45 grid-points.

\begin{table}
	\centering
	\caption{Danger category changes after applying Geo-Ind ($r=300$ meters) in Fukushima.}
	\label{table:fuku-geoind300}
	\resizebox{\columnwidth}{!}{
		\begin{tabular}{lrrrrrr}
			\textbf{Geo-Ind: 300m} &     \textbf{1} &     \textbf{2} &     \textbf{3} &     \textbf{4} &     \textbf{5} &      \textbf{Number} \\
			\cmidrule[0.25pt]{1-1}
			\textbf{Original} &       &       &       &       &       &   \textbf{of points}       \\
			\midrule
			\textbf{1}        & 79.7\% & 19.3\% & 1\% & 0.003\% & 0.001\% &  1,354,110 \\
			\textbf{2}        & 41.5\% & 49.5\% & 9\% & 0.023\% &0.01\% &   650,486 \\
			\textbf{3}        & 8.7\% & 35.9\% & 52.2\% & 2.3\% & 0.9\% &   229,848 \\
			\textbf{4}        & 2.5\% & 3.3\% & 49.3\% & 29.8\% & 15.1\% &    10,489 \\
			\textbf{5}        & 3.9\% & 1.7\% & 34.7\% & 29.3\% & 30.4\% &     5,067 \\
		\end{tabular}
	}
\end{table}

\begin{table}
	\centering
	\caption{Danger category changes after applying the Rounding mechanism (2 decimals) in Fukushima.}
	\label{table:fuku-round2}
	\resizebox{\columnwidth}{!}{
		
		\begin{tabular}{lrrrrrr}
			\textbf{Rounding: 2} &     \textbf{1} &     \textbf{2} &     \textbf{3} &     \textbf{4} &     \textbf{5} &      \textbf{Number} \\
			\cmidrule[0.25pt]{1-1}
			\textbf{Original} &       &       &       &       &       &    \textbf{of points}      \\
			\midrule
			\textbf{1}        & 89.3\% & 10.3\% & 0.3\% & - & 0.001\% &  1,354,110 \\
			\textbf{2}        & 30.2\% & 64\% & 5.8\% & 0.003\% &   -    &   650,486 \\
			\textbf{3}        & 0.7\% & 22.6\% & 74.8\% & 1.6\% & 0.3\% &   229,847 \\
			\textbf{4}        & 0.2\% & 0.01\% & 43.3\% & 39.6\% & 16.9\% &    10,490 \\
			\textbf{5}        &   0.9\% &   - & 9.3\% & 42.1\% & 47.6\% &     5,067 \\
		\end{tabular}
	}
\end{table}

\begin{figure}
	\centering
	\includegraphics[width=0.8\columnwidth]{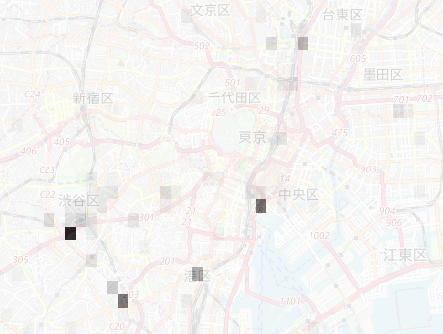}
	\caption{Prior probability of visiting locations in Tokyo (white - low probability, black - high probability).}
	\label{figure:heatmap-tokyo}{}
\end{figure}

\smallskip\noindent\textbf{\textit{Why optimal remapping does not work for \ac{MCS}.}}
Even though GeoInd-OR was designed to increase utility while preserving privacy, we observe that, in the \ac{MCS} case, utility roughly stays the same (Figure \ref{fig:utility-tokyodif}), and privacy slightly increases, both in decreasing the number of vulnerable users and in increasing the spatial gain.
The reason for this mismatch is that this mechanism was designed in the context of \acp{LBS}, where remapping locations to places where the user is likely to be is bound to provide good utility on average. 
However, in Safecast, the utility does not depend on the locations themselves, but on the associated measurements. Remapping the location, however, concentrates measurements in these popular locations, effectively polluting the measurements. We illustrate this effect in Figure \ref{figure:heatmap-tokyo}, which represents the prior probability of users' locations over all locations in Tokyo (low in white, high in black). In the low probability areas, most locations have the same probability, thus remapping has a randomizing effect. However, when there is a location with high probability, all locations are remapped to this popular location. We note that, while significantly hurting utility, this effect creates artificial clusters that reduce the adversary's precision and recall, thus increasing privacy.

\smallskip\noindent\textbf{\textit{The case of high precision measurements.}}
Safecast also uses the crowdsourced measurements to monitor radiation
\emph{hotspots} that could be dangerous for public health. For this case, location precision is highly
important, both to understand the dangers it can cause and to keep 
low costs if experts have to be sent to study the origin of the abnormality. 

\begin{figure}
	\centering
	\includegraphics[width=0.95\columnwidth]{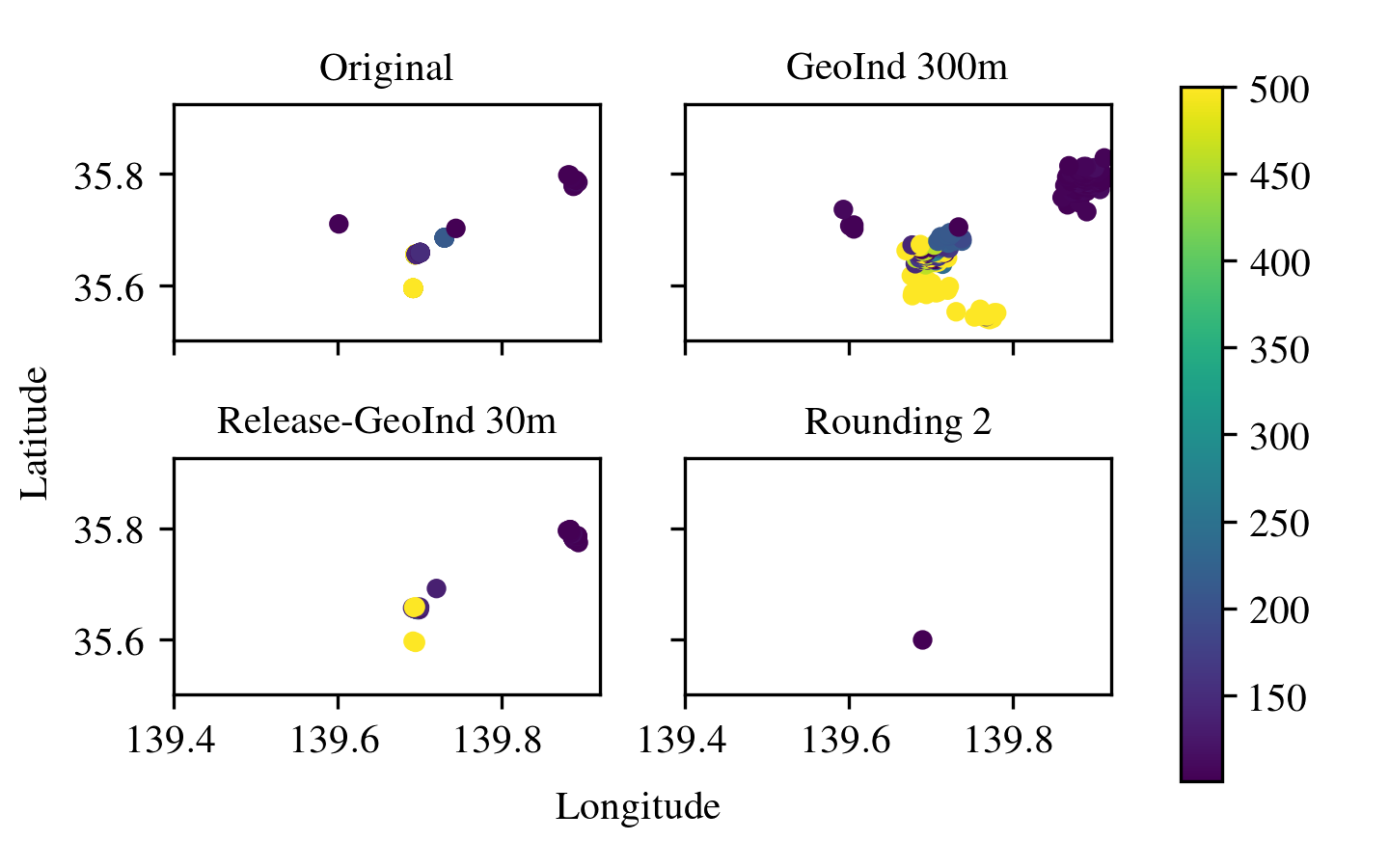}
	\caption{Safecast: Hotspot detection for areas with at least 100 cpm. Comparison of various defenses vs the original hotspots.}
	\label{figure:hotspots}
\end{figure}

We study the impact of \acp{LPPM} on hotspot localization by looking for
locations with more than 100 cpm radiation after averaging the measurements over the last 270 days 
but \textit{before interpolating the data}. This is to avoid that interpolation modifies the
position of the hotspots, or even eliminates them.
We show the results of detection when using the raw measurements (top left), and after the application of Release-GeoInd 30m (bottom left), GeoInd 300m (top right), and Rounding 2 (bottom right) in Figure \ref{figure:hotspots}.
We see that noise-based mechanisms spread the measurements and, as the noise increases, create additional hotspots.
Thus, these mechanisms are useless for hotspot detection: the results cannot be properly interpreted.
Imagine a hotspot in a place known to present high radiation, thus being already closely monitored by the authorities. Finding such hotspot is not alarming. However, after spreading, the finding of hotspots conveys a much different message, especially when they appear in zones that had low radiation in the past.

Generalization such as Rounding 2, which provides a good privacy-utility tradeoff for aggregated statistics, also performs poorly. In this case, the defense causes hotspots to disappear, potentially causing a dangerous situation if a high radiation zone is marked as safe. We also carry out experiments with hiding mechanisms and find that, similarly to Rounding, they miss some of the original hotspots.

\smallskip\noindent\textbf{\textit{Safecast takeaways.}} Considering only the privacy loss, GeoInd variants (except GeoInd 50m) and Rounding to 2 decimals seem to offer the best performance, while Random sampling and Release's protection is generally bad in terms of precision, and also too dependent on users' movement profiles. 
However, an analysis of the utility impact indicates that \textit{none of the existent \acp{LPPM} is well suited} for the Safecast setting. The semantic interpretation results indicate that even if two defenses produce similar average results, the outliers they create can convey opposite messages. Furthermore, even a slight addition of noise or generalization can hinder the project's ability to correctly locate abnormal events. These limitations effectively prevent Safecast from deploying them to protect their users' privacy.\footnote{This statement was verified in communication with Safecast.}

\subsubsection{Radiocells}
\begin{figure}
	\centering
	\includegraphics[width=0.9\columnwidth]{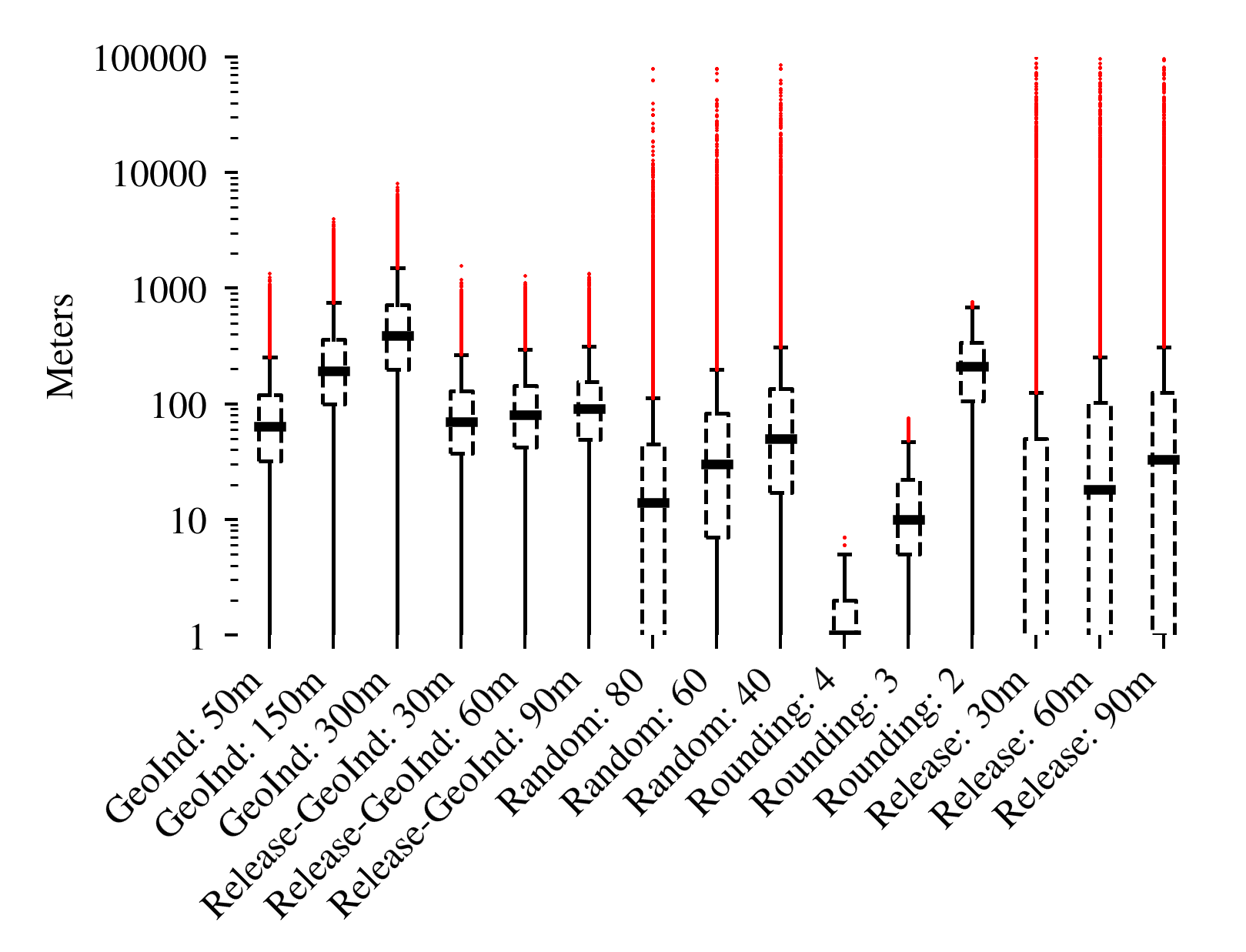}
	\caption{Radiocells: Utility loss (distance to tower location).}
	\label{figure:utility-radio}
\end{figure}

Radiocells' utility function is rather different than the one for Safecast. Instead of averaging measurements associated to a location, Radiocells averages all reported coordinates associated to an antenna to derive its position. We show the related utility loss for different \acp{LPPM} in Figure \ref{figure:utility-radio}. 

All \ac{GeoInd} variants induce high utility loss, with medians between 80 and 400 meters, and with outliers beyond 2 kilometers.
Surprisingly, in this use case hiding mechanisms (Release and Random) have many outliers. After manual inspection, we found out that several users had inconsistent measurements. For instance, a user was swapping her measurements' longitudes and latitudes in a random pattern. Other outliers are caused by providers moving their antennas IDs creating mixed measurements for a given ID.
Furthermore, hiding defenses also influence the number of antennas located. In our dataset, we detect from 10.2\% up to 18.6\% fewer antennas when the Release defense is used, and the Random mechanism eliminates from 2.6\% up to 13.7\% of them.

The best mechanism in the Radiocells dataset is Release GeoInd which offers on average lower utility loss than other \acp{LPPM} and provides acceptable privacy. However, some antennas might be moved over a kilometer away. The next best alternative is Rounding 2 that has a higher median utility loss but no  outliers. 
However, as the goal of the project is to \textit{accurately} detect antennas in order to give individuals the ability to geolocate themselves offline or to enable scientific studies, a median error of 100 meters (Release GeoInd) or 200 meters (Rounding 2) is considered too large and precludes Radiocells from deploying them.

%% file: threat.tex

\subsection{Validating the Inference Strategy}
\label{sec:attacks}
 
We now validate the suitability of DBSCAN as strategy for inferring areas and points of interest in the context of MCS. Specifically, we test its suitability to identify workplaces on data from Safecast and OpenStreetMaps. As both projects' public data contain identifiable information about their users, we can validate the inferences against information available on other online platforms. We choose workplaces for ease of validation, but we note that it is just one of many inferences that could be done using location data~\cite{drakonakis2019please}. Our results below confirm that DBSCAN is a suitable choice as basis to compute areas and POIs to input in our privacy metrics.

\smallskip\noindent\textbf{Ethical considerations.} For these experiments we do not collect any personal data other than that made publicly available by the MCS projects. We have limited our inferences to the minimum to validate the suitability of DBSCAN. We only report aggregated or anonymized data such that no individual's data is exposed. We have notified the service providers about our findings, and we have shared our code with them so that they can make informed decisions regarding improvements of the privacy situation. 
Our code is open-source so that it can also be used by other crowdsourcing applications and improved by the research community~\cite{code}. 
This procedure has been approved by EPFL's Human Research Ethics Committee (HREC).

\begin{table}
	\centering
	\caption{Safecast dataset statistics.}
	\label{table:safecast-api}
	\begin{tabular*}{\columnwidth}{llll}
		\textbf{Measurements} & \textbf{Users} & \textbf{Avg measurements}  & \textbf{Avg days} \\ \midrule
		\textless{}10k  & 213   & 3,331   & 5    \\ 
		10k-100k  & 230   & 38,341  & 20    \\ 
		100k-1M & 87    & 270,387  & 105  \\ 
		\textgreater{}1M & 10    & 1,958,760   & 632   \\ 
	\end{tabular*}
\end{table}

\smallskip\noindent\textbf{Safecast.}
To evaluate the effectiveness of DBSCAN in different situations, we split the users in the dataset into four groups according to their amount of measurements they report. For each group, Table \ref{table:safecast-api} shows the number of users, their average amount of measurements, and the average number of days in which they took at least one measurement. From each group we select as targets for inference the 10 users with the most measurements that provide their real names. Since in the group with \textgreater{}1M there are only 4 users with real names, we end up with 34 target users in total. This allows us to manually validate our inferences in reasonable time.

\smallskip\noindent\textit{Identifying workplaces.} 
\label{sec:empiricalattacks} 
We run DBSCAN on every users' measurements during working hours (Monday to Friday from 9AM to 5PM). We configure DBSCAN to find clusters with at least 80 points separated by 60 meters and, if no clusters are found, we increase the distance by 30 meters (up to 120 meters maximum) and decrease the number of points by 15 (down to 35 points). These parameters have been chosen empirically to optimize the adversary's success, see Appendix~\ref{comparison-privacy-gain}. To keep the manual analysis feasible, we only consider the five clusters with the highest number of points.

We expect that the users' workplace is one of the POIs within the inferred geographic area. In many cases, however, this area is large and contains many \acp{POI}. 
To ease manual validation, we use X-means clustering~\cite{Pelleg00x-means:extending} to split these large clusters, and consider as \acp{POI} the centroids of the two largest subclusters. We end up with at most 10 \acp{POI} per user. We use the MapQuest API~\cite{mapquest} to obtain these locations' addresses and, if existing, the names of the businesses at those coordinates. 
We recall that in our LPPM evaluation below, we consider all points in the clusters as input for the POI Gain metric. This represents a resourceful adversary that can afford checking manually all the points and filter out those corresponding to businesses. We note that considering more points could cause more false positives, but the semantics of locations often makes it easy to filter these out, e.g., lakes or parks can be usually discarded as workplaces.

Once we have candidate workplaces, we validate them using social networks such as Twitter or LinkedIn, or the users' personal webpages. We note that 9 of our target users did not have a publicly available profile or had too common names to find their correct information, thus we could not validate their inferred workplaces.
Overall, we recover the workplace of 35\% of the target users. This result is consistent across the groups: 40\% of the users with less than 10k measurements, 20\% of the users with 10k-100k measurements, 50\% of the users with 100k-1M measurements, and 25\% of the users with more than 1M measurements. We conclude that DBSCAN performs well for POI identification irrespectively of the amount of data shared by the users. Surprisingly, this means that privacy seems not correlated to the volume of data made available to the adversary. On the contrary, it seems to be highly dependent on the collection patterns of the users. We observe that people fall in one of two categories: (i) Those who travel to different places with the goal of obtaining measurements, whose work addresses cannot be inferred; and (ii) those who measure radiation during their daily activities, whose work place we can find. The Safecast co-founders, who are the top contributors in terms of data points, fall in the first category, explaining the lower inference power for users with more than 1M measurements.

Our results confirm recent findings in the literature regarding personal information inferences from location data\cite{Efstathiades,drakonakis2019please}. Yet, we want to stress that the threat may be worse for \ac{MCS}, due to the volume of data exposed by participants. For reference, Safecast's lowest contributing group has on average 3k measurements per user (see Table \ref{table:safecast-api}) while in the Twitter analysis performed by Drakonakis et al.~\cite{drakonakis2019please} only the top contributing users (less than 0.06\%) have more than 3k geolocated tweets. Thus, even if the number of \ac{MCS} users is not as large as social networks' users, we expect a significant fraction of them to be vulnerable to privacy attacks.

\smallskip\noindent\textit{Other \acp{POI}.}
A deeper analysis of the times and semantics of the \acp{POI} identified by DBSCAN revealed further information about Safecast users. 
Among others, we could infer two users' membership to specific organizations: one member of the Scientology church who reported many points from the Church of Scientology Celebrity Centre in a major city; and a Masonic lodge member who regularly visited the lodge headquarters. We could verify this information online for both users.
We also identified two work-related activities: a US-based scientist working on a project about radiation around a lake in the Southern part of the US, and a photographer working in a Japanese city. We validated these inferences using Research Gate and the webpage of the artist, respectively. 
Finally, we could follow the education steps of a European PhD student. Her points of interest over time reveal the university where she obtained her master's degree, an exchange with another European university, and the university where she is completing her doctoral studies. We verified these facts on her CV available online.

\smallskip\noindent\textbf{OpenStreetMaps.}
\label{sec:empiricalattacks_osm}
Contrary to Safecast, OSM does not have an open API for accessing users' data. Yet, traces from users who have chosen to make their data available can be easily obtained from OSM's website.\footnote{\url{https://www.openstreetmap.org/traces}} To minimize the impact on OSM servers, and comply with their non-crawling policy, we manually downloaded data for 30 users with a large amount of contributions,\footnote{\url{http://resultmaps.neis-one.org/oooc}} of which 17 used their actual names (or indicative nicknames). 
Although the majority of the points in the dataset were rather old (most of them at least 7-8 years old), we were able to verify previous workplaces for 3 of the 17 users (17\%). We note that, for some users, we found out that they did not have a standard place of employment during data collection period (e.g., students). However, for \textit{all} users, their \acp{POI} were within the area where they worked or lived. We used this fact to infer two of the users' short vacation trips which we manually verified with information publicly accessible from their social media accounts. 

%% file: Framework.tex

In this section, we elaborate on technical and non-technical steps to enhance privacy at smaller utility cost in the context of MCS applications.

\subsection{Towards Effective Defenses}
 
We first discuss possible strategies to improve the trade-off between users' privacy and MCS utility.

An unexplored approach is the use of advanced cryptographic protocols to compute the values of interest for MCS without revealing the users'
individual values to the providers \cite{demmler2015aby}. 
For instance, users could use multi-party computation to collaboratively compute aggregates and only report the result to the provider.   
However, cryptographic approaches require high computational power on the users' side and increase the bandwidth needs to perform the joint computation.
Furthermore, this would limit the availability of raw measurements for analysis other than those predefined by the cryptographic protocols,
which is at odds with the principles of open data and open science defended by most of the MCS platforms.

In our evaluation, we only considered spatial generalization. Another avenue to explore would be to also generalize the time dimension. On its own time
obfuscation cannot hide patterns revealed by repeated visits. However, combined with full de-identification and hiding of users could reduce the inference power of the adversary. For instance, the MCS service provider could release a batch of measurements once a day or once a week without linking these to any user identifier.
These techniques would be cheaper than the use of cryptography, but require trust on the service provider to properly apply sanitization and protect the raw data. 

A third research path is the co-design of defenses and aggregation algorithms. In this paper, we have considered that the output of the LPPMs is directly input to the utility functions currently used by MCS providers. However, it would be possible that the providers adapt their data processing to account for noise, using statistical methods or machine learning, as done in fields that rely on noisy sensors~\cite{shi2010prisense} or train in different settings from which they are deployed~\cite{DurrettKBPAMLP17,OverdorfG16,tuia2016domain}.

Finally, MCS could provide users with dedicated local software (e.g., building on our evaluation method) to alert them regarding the privacy dangers of publishing raw location data. Such a system would allow them to selectively hide some of their measurements, reducing the confidence of inference attacks.
We note that, when building such a tool, one would like to consider attacks beyond the POI-based inferences considered in this paper. For instance, it has been shown that co-locations can unveil social links~\cite{crandall2010inferring,eagle2009inferring}. We run a preliminary evaluation to learn whether our MCS setting is also prone to such an attack.
We identified 50 unique pairs of users with real names and at least one co-location (similar latitude, longitude, and time) in the Safecast dataset.  
We could validate 16 of these pairs as real friendships using information available on online social networks, i.e., yielding a 32\% correct inference rate. 
Note that many of the other pairs could not be verified because either users were not part of any social network or they did not publicly reveal their social links. More advanced methods, such as measuring the amount of time two users are co-located or the number of different locations where two users jointly report their locations~\cite{crandall2010inferring,wang2014pgt,backes2017walk2friends} could further improve these results. Therefore, new defenses need to also obfuscate co-locations~\cite{olteanu2019colocation}.

\subsection{Privacy Considerations for Developers}
In our study, we identified a number of issues related to the collection and sharing of data that, 
even though cannot fully prevent inference, could make inference attacks detectable and could render potential attackers accountable. 

A first consideration to make is the type of policy under which \ac{MCS} publish the collected data. While making large datasets available to everyone for unrestricted use is admirable, and certainly of high value for the academic community, it can have serious implications for the altruistic contributors. To reduce this risk, developers could add clauses to the policies that not only mandate that use of the data is properly acknowledged, but also that it is well documented, implying that researchers or other individuals have to disclose how they have processed the data, and for which purpose.

Second, both Safecast and Radiocells datasets are available for download without the need for authentication. This hinders traceability of who has the data, and thus enables stealthy attacks where nor the users neither the applications are aware of the danger. Like in other projects that make data available for research and other purposes (e.g., the Drebin project\footnote{\url{https://www.sec.cs.tu-bs.de/~danarp/drebin/}}), these sites could require simple registration to maintain a log of who has had access to the datasets. Together with the previous requirement, which would include documentation of sharing, it should help mitigate the risks.

Third, these applications typically do not perform any control on who are the contributors. This poses a particular problem when it comes to children. In many jurisdictions, children's data are subject to particular legislation~\cite{eu:gdpr,coppa}, and in particular require the parents' consent to be collected and processed. The lack of control upon collection implies that the datasets could contain children's geo-located data collected illegally. Adding control would solve this problem and also support the previous two points.

Finally, the datasets we studied contain data from users from all over the world. These users, therefore, are subject to different legislations that regulate how their data can be processed. While this may not be a problem for corporations or criminals that want to exploit the datasets, it creates a hurdle for researchers who have to obtain approval from their institution for data processing. This problem arised during our discussions with our institution's Ethical Review Committee, and almost caused us to stop the project. In other words, lack of proper documentation may limit the free use of the data for science, effectively hindering one of the main goals of these applications. Better documentation as to the origin of data and its use possibilities would greatly facilitate the process.

%% file: RelatedWork.tex
We have covered the related work on \acp{LPPM} in Section \ref{sec:defs} and the previous work on privacy quantification in Section \ref{sec:metrics}.
We complete this review of the literature with previous research on human mobility and its privacy implications.

Similar to \cite{liao2006location,liao2007learning,heo2012real,cho2016exploiting,mathew2012predicting,krumm2007inference,hoh2006enhancing,freudiger2011evaluating,urnerassessing}, our \acp{POI} extraction attack is based on machine learning. 
Gambs and Killijian \cite{gambs2014anonymization} also rely on \acp{POI} inference to build mobility Markov chains and de-anonymize traces.
Gonzalez et~al.~\cite{gonzalez2008understanding} and Song et~al.~\cite{song2010limits} study anonymized mobile phone data. Their results indicate that human trajectories have a high degree of temporal and spatial regularity, and that an individual's location data history is a unique identifier. De Montjoye et~al.~\cite{de2013unique} investigate how the uniqueness of mobility traces decays depending on their resolution. They show that uniqueness cannot be avoided by lowering the resolution of a dataset. 
While these works aim at understanding the uniqueness of individuals or de-anonymize them, we focused on inferences that rely on labeled traces.

Similar to us, Gambs et~al.~\cite{gambs2010show} develop a platform for evaluating various sanitization methods and attacks on geo-located data. They focused on evaluating \acp{LBS}, while we evaluate the effectiveness of defenses on \ac{MCS} applications. We also use different privacy metrics, and utility functions, tailored to the \ac{MCS} scenario.  
Finally, Drakonakis et~al.~\cite{drakonakis2019please} explore the privacy loss stemming from by public location metadata. 
They propose a tool to infer users' regions of interest and, by experimenting with data gathered from Twitter, they illustrate the accuracy of their tool in pinpointing  users' sensitive locations. Furthermore, they highlight how the spatial data provide additional context on the information shared by the user. We use similar techniques to prove that these inferences are also possible in \ac{MCS}. 
For further information about the security and privacy landscape of location data we refer the reader to the surveys in \cite{terrovitis2011privacy,Krumm09,ghinita2013privacy, primault2018long}.

%% file: Conclusion.tex

Mobile crowdsourcing is an increasingly popular way to collect geo-located data from millions of contributors.
We present the first study on privacy implications of \ac{MCS} applications.
We study the applicability of well-established location privacy
defenses created for \acp{LBS}. We show that neither the location privacy and utility metrics typically found in the literature
nor the existing privacy-preserving mechanisms are well-suited for the \ac{MCS} case. On the one hand, given the
persistent patterns stemming from continuous collection, these solutions provide less privacy than in the 
case of \acp{LBS} where locations are revealed once. Second, the existing mechanisms are optimized to
provide utility regarding the location of the users, but \ac{MCS} applications rely on measurements associated 
to these locations, or on some function of the locations. Therefore, state-of-the-art defenses have 
a detrimental impact on the \ac{MCS} utility. 

In conclusion, we identify an underexplored space in the location privacy literature, that is of
practical relevance for many new applications. We have outlined promising lines to 
improve the situation. We hope that our findings spawn new research
that soon enables the deployment of privacy-preserving crowdsourcing applications.

%% file: Appendix.tex
\subsection{Density Based Clustering (DBSCAN)}
\label{ap:dbscan}

The algorithm receives as input all locations (also referred to as points) reported by a user, the minimum required amount of points per cluster, and the maximum allowed distance between the cluster's points. 
It outputs a label for every point, indicating to which cluster it belongs, or if it has been labeled as noise.

DBSCAN starts by randomly selecting a point $c$. Then, it finds all points $p$ that are in distance $\epsilon$ from this point. Then, from the points $p$ reachable from the first point, it tries to find more points $q$ where $q$ are reachable directly from $p$ but not from $c$. If at the end of this procedure the minimum points have not been reached, it moves to another random point and starts all over again. In order to use our locations which are in latitudes and longitude, we converted the distance $\epsilon$ to radians first. Moreover, we used a ball tree data structure to speed up the neighbors queries.

\subsection{Geo-Indistinguishability}
\label{ap:geoind}
The  noise is drawn by first transforming the location to polar coordinates. Then, the angle is drawn randomly between $0$ and $2\pi$  while the distance is drawn from
			$$C^{-1}(\rho)= -\frac{1}{\epsilon}(W^{-1}( \frac{\rho-1}{e} )+1) $$
 with $W^{-1}$  denoting the $-1$ branch of the  Lambert\,W function. Finally, the generated distance and angle are added to the original location.

\subsection{Optimal Remapping}
\label{ap:OR}
For the optimal remapping technique we follow these steps;
For performance reasons, we first round each location to 3 digits, in order to merge nearby locations together.
Then, we calculate the  probability of each coordinate. 
Afterwards, we convert all coordinates to a Cartesian system using their distance from the center of the Earth.
A useful tutorial on this can be found in \cite{earthpy}. 
Using the Cartesian coordinates we build a KD-Tree for efficient nearest neighbor calculations.
Then, for every location where GeoInd has been applied, we query all nearest neighbors in a region $r'$.
This  $r'$ is set to be as the 99\% percentile of the distribution that generated the parameter $r$ used in GeoInd.
In other words, the user has 99\% chance of being remapped somewhere within this distance. 
For all neighboring points, we compute the posterior and then, we calculate the geometric median of those coordinates using the iterative Weiszfeld's algorithm. 
The geometric median minimizes the average Euclidean distance and hence, returning us the new, optimal (in terms of utility as privacy should remain the same) location.

\subsection{Defenses evaluation}
 \label{DefensesSubsectionAppendix}

\begin{figure}
	\centering
	\includegraphics[ scale=0.9]{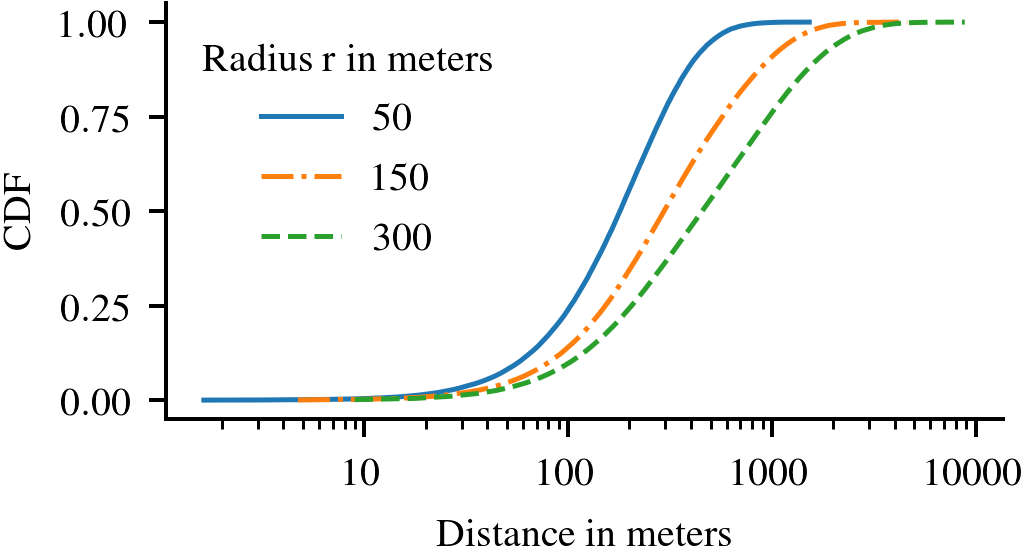}
	\caption{\ac{GeoInd} noise magnitude for different radius ($l=\ln(1.6)$).}
	\label{figure:geoind-cdf}
\end{figure}

\begin{figure}
	\centering
	\includegraphics[width=0.95\columnwidth, scale=0.9]{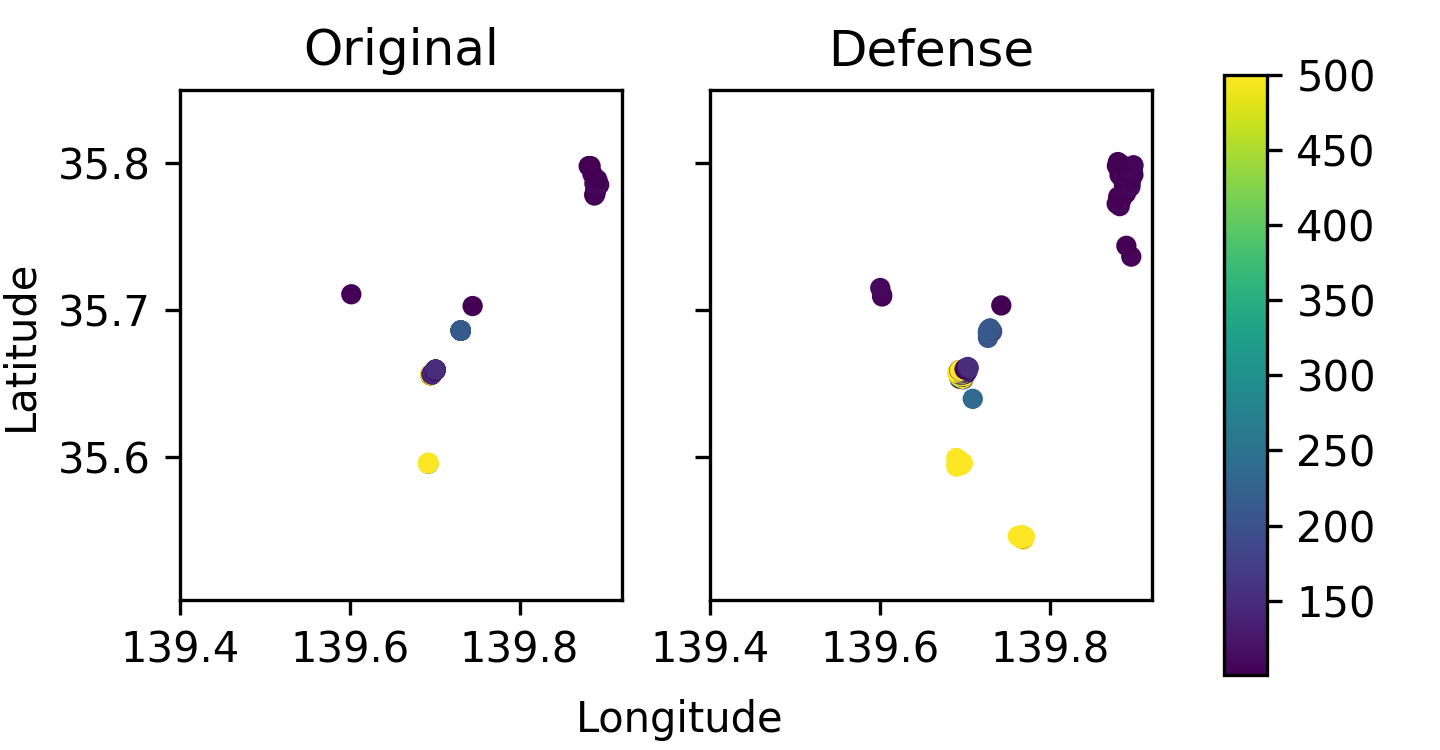}
	\caption{Safecast: Hotspot detection for areas with at least 100 cpm. The presented defense is \ac{GeoInd} with 50m parameter.}
	\label{figure:hotspots-geoind50}
\end{figure}

\begin{figure*}
\centering
\includegraphics[scale=0.65]{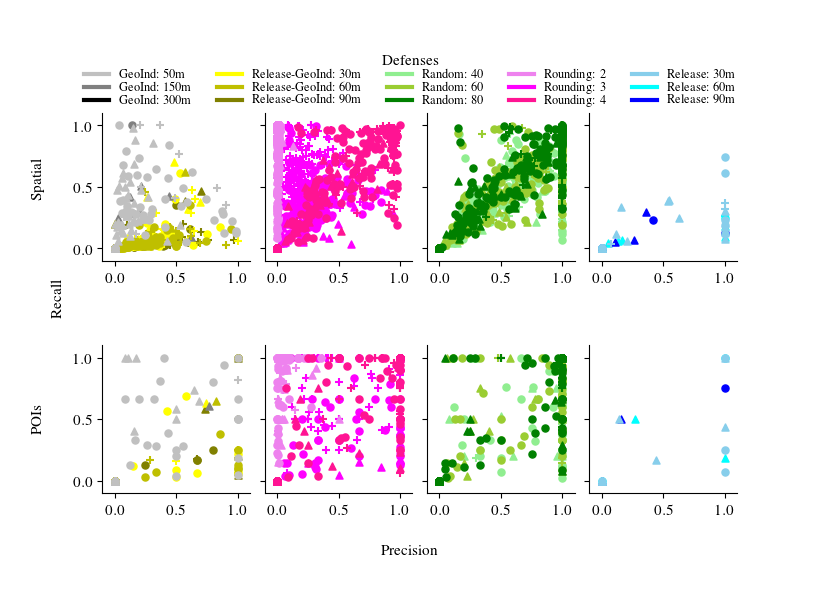}
\caption{Safecast: Privacy gain for each of the defense mechanism (whole dataset).}
\label{figure:world-geoind}
\end{figure*}

We include three more figures to complement the defense evaluation:
\begin{itemize}[nosep]
\item Figure \ref{figure:geoind-cdf}  portrays the CDF of the noise added by GeoInd. This noise is added on all GeoInd variants (Optimal Remapping and Release-GeoInd) and it is controlled by either the radius ($r$) or the privacy parameter ($l$).
\item Figure \ref{figure:hotspots-geoind50} illustrates the hotspot detection results when GeoInd 50m is used. Even a slight addition of noise spreads the locations, not allowing Safecast to accurately detect elevated radiation regions.
\item In Figure~\ref{figure:world-geoind} we present the privacy gain results for each of the defense mechanisms for the whole Safecast dataset. 
\end{itemize}

\subsection{Experimental results}
\label{Exp_res_appendix}

\begin{figure}
	\centering
	\includegraphics[width=0.8\columnwidth]{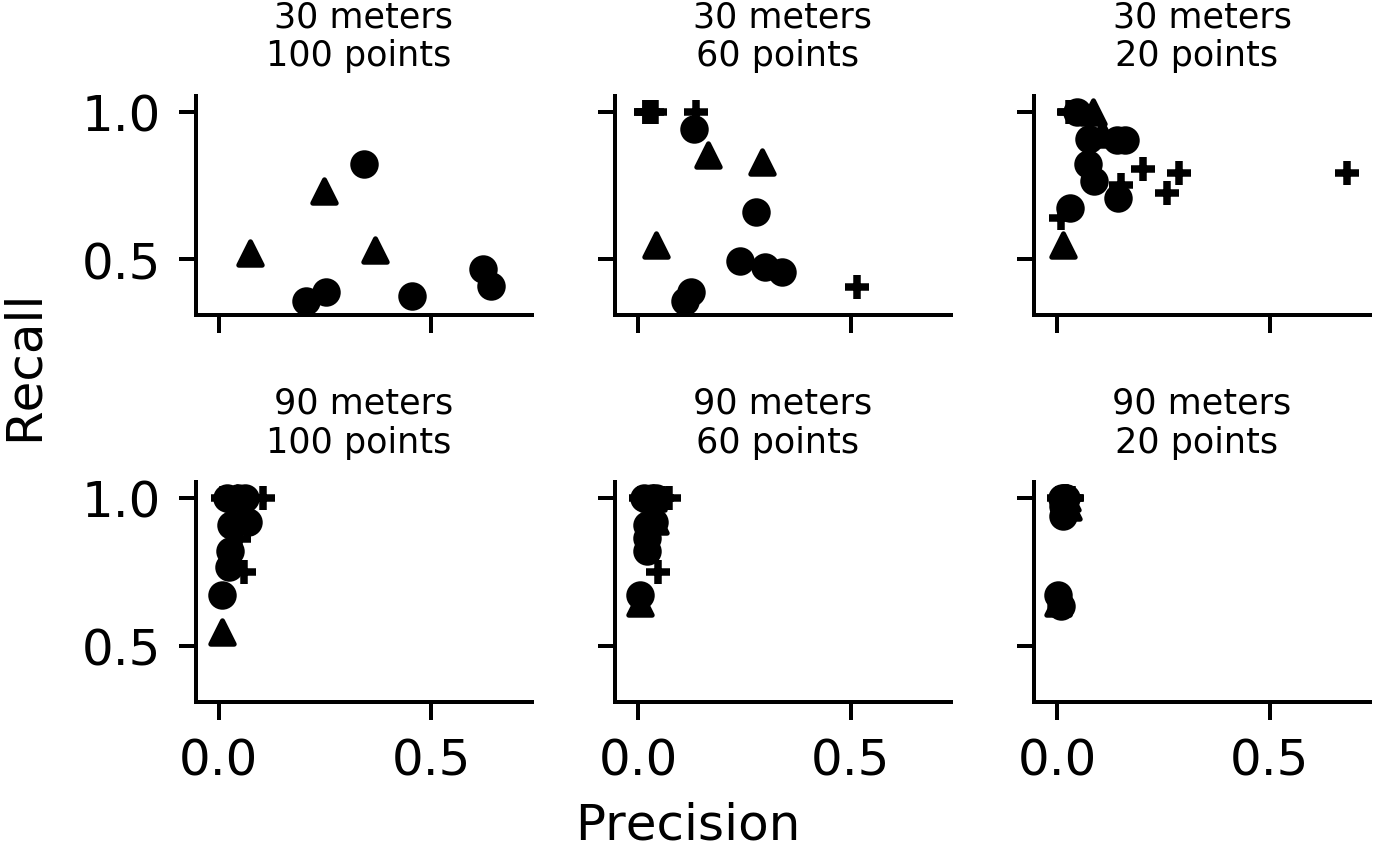}
	\caption{Precision and recall vs. clustering parameters for \ac{GeoInd} ($r=50$m) in Tokyo.}
	\label{figure:geoind-clustering}
\end{figure}

\begin{figure}
	\centering
	\includegraphics[width=\columnwidth,scale=0.9]{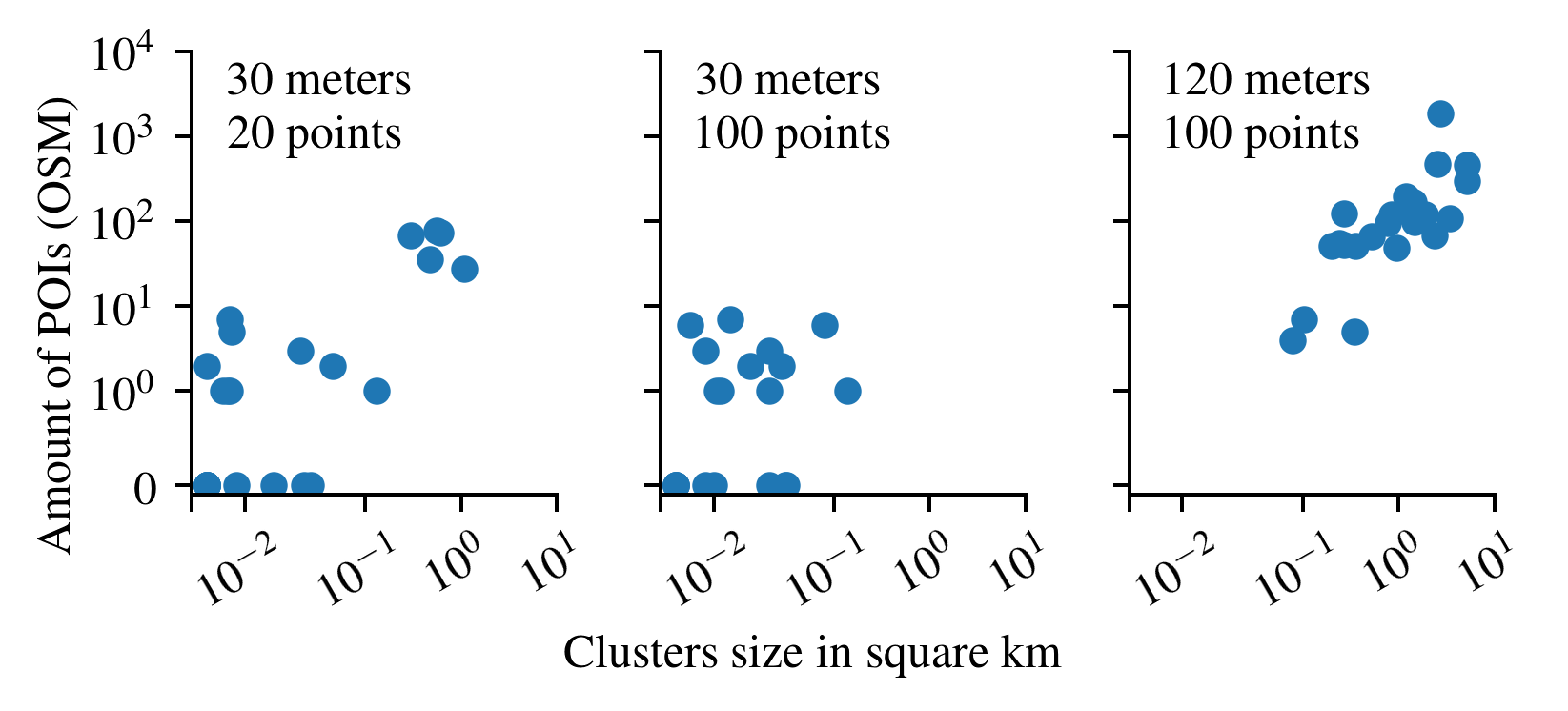}
	\caption{Clusters' size and amount of \acp{POI} per cluster vs. clustering parameters with \ac{GeoInd} ($r=50$m) in Tokyo.}
	\label{figure:geoind-privacy}
\end{figure}

\smallskip\noindent\textbf{\textit{Adjusting the clustering parameters.}}
\label{comparison-privacy-gain}
We now study the influence of the DBSCAN clustering parameters on our results. We show the difference in precision and recall for \ac{GeoInd} ($r$=50 meters) when we vary both the maximum distance and the minimum number of point per cluster in Figure \ref{figure:geoind-clustering}.
As we increase the maximum distance between points and decrease the minimum required points per cluster, the results concentrate on the upper left corner of the diagram.
This is because as the parameters become `looser', the resulting clusters grow in size increasing recall (more likelihood of covering all users' original clusters) but reducing precision due to many false positives. Furthermore, increasing the cluster size increases the adversary's cost, as the clusters contain a larger number of \acp{POI} (Figure \ref{figure:geoind-privacy}) which requires more filtering and increases the probability of having false positives.